\documentclass[a4paper,11pt,twoside]{article}
\makeatletter

\usepackage{amsmath,amssymb}

\usepackage[english]{babel}
\usepackage{polski}
\usepackage{cancel}
\usepackage{color}
\usepackage{enumitem}
\usepackage{graphicx}
\usepackage{geometry}

\usepackage{hyperref}
\usepackage{cite}
\usepackage{url}

\usepackage[amsmath,hyperref,thmmarks]{ntheorem}

\numberwithin{equation}{section}
\allowdisplaybreaks[1]
\newcommand{\be}{\begin{equation}}
\newcommand{\ee}{\end{equation}}

\newcommand\bbone{{\mathbb{I}}}

\newcommand{\labitem}[2]{\def\@itemlabel{\textbf{#1}}\item\def\@currentlabel{#1}\label{#2}}

\theoremsymbol{}
\theorembodyfont{\slshape}
\theoremheaderfont{\normalfont\bfseries}
\theoremseparator{}

\theorembodyfont{\upshape}
\theoremsymbol{\ensuremath{\blacklozenge}}

\theoremstyle{nonumberplain}
\theoremheaderfont{\scshape}
\theorembodyfont{\normalfont}
\theoremsymbol{\ensuremath{\blacksquare}}

\qedsymbol{\ensuremath{_\blacksquare}}
\theoremclass{LaTeX}

\newcommand{\institute}[1]{\newcommand{\@institute}{#1}}

\renewcommand{\maketitle}{
\vspace*{0.5\baselineskip}
{
\center\LARGE\noindent\@title\par
}%
\vspace{1.5\baselineskip}
{
\center\normalsize\noindent\ignorespaces\@author\par
}%
\vspace{0.5\baselineskip}
{
\center\normalsize\ignorespaces\@institute\par
}%
\vspace{2\baselineskip}
}%

\renewenvironment{thebibliography}[1]{%
\section*{References}%
\frenchspacing\small%
\begin{list}{[\arabic{enumi}]}%
{%
\usecounter{enumi}\parsep=2pt\topsep 0pt%
\settowidth{\labelwidth}{[#1]}%
\leftmargin=\labelwidth\advance\leftmargin\labelsep%
\rightmargin=0pt\itemsep=1pt\sloppy%
}%
}{\end{list}}

\begin{document}

\title{$\kappa$-Poincar\'e invariant quantum field theories\\
with KMS weight}

\author{Timoth\'e Poulain, Jean-Christophe Wallet}

\institute{%

\textit{Laboratoire de Physique Th\'eorique, B\^at.\ 210\\
CNRS and Universit\'e Paris-Sud 11,  91405 Orsay Cedex, France}\\
e-mail: \href{mailto:timothe.poulain@th.u-psud.fr}{\texttt{timothe.poulain@th.u-psud.fr}}, \href{mailto:jean-christophe.wallet@th.u-psud.fr}{\texttt{jean-christophe.wallet@th.u-psud.fr}}\\[1ex]%
}%

\date{\today}

\maketitle


\begin{abstract} 
A natural star product for 4-d $\kappa$-Minkowski space is used to investigate various classes of $\kappa$-Poincar\'e invariant scalar field theories with quartic interactions whose commutative limit coincides with the usual $\phi^4$ theory. $\kappa$-Poincar\'e invariance forces the integral involved  in the actions to be a twisted trace, thus defining a KMS weight for the non-commutative (C*-)algebra modeling the $\kappa$-Minkowski space. In all the field theories, the twist generates different planar one-loop contributions to the 2-point function which are at most UV linearly diverging. Some of these theories are free of UV/IR mixing. In the others, UV/IR mixing shows up in non-planar contributions to the 2-point function at exceptional zero external momenta while staying finite at non-zero external momenta. These results are discussed together with the possibility for the KMS weight relative to the quantum space algebra to trigger the appearance of KMS state on the algebra of observables.
\end{abstract}
\vskip 1 true cm

\newpage

\section{Introduction.}
It is widely believed that the classical notion of space-time is no longer adequate at the Planck scale to reconcile gravity with quantum mechanics. One possible attempt to reach this goal comprises to trade the continuous smooth manifold describing the space-time by a non-commutative (quantum) space \cite{Doplich1}. In this spirit, the $\kappa$-Minkowski space-time appears  in the physics literature to be one of the most studied non-commutative spaces with Lie algebra type non-commutativity and is sometimes regarded as a good candidate for a quantum space-time to be involved in a description of quantum gravity at least in some limit. Informally, it may be viewed as the enveloping algebra of the Lie algebra $[x_0,x_i]=i\kappa^{-1} x_i,\ [x_i,x_j]=0,\ i,j=1,\cdots, d$, where the deformation parameter $\kappa$ has dimension of a mass. The $\kappa$-Minkowski space-time has been characterized a long time ago in \cite{majid-ruegg} by exhibiting the Hopf algebra bicrossproduct structure of the $\kappa$-Poincar\'e quantum algebra \cite{luk1} which (co-)acts covariantly on it and may be viewed as describing its quantum symmetries. A considerable amount of literature has been devoted to the exploration of algebraic aspects related to $\kappa$-Minkowski space and $\kappa$-Poincar\'e algebra, in particular dealing with concepts inherited from quantum groups \cite{leningrad} as well as (twists) deformations. For a comprehensive recent review, on these algebraic developments, see e.g \cite{luk2} and the references therein. Besides, the possibility to have testable/observable consequences from related phenomenological models has raised a growing interest and resulted in many works dealing for instance with Doubly Special Relativity together with modified dispersion relations and relative locality \cite{ame-ca1,reloc}.\bigskip

Once the non-commutative nature of the space-time is assumed, Non-Commutative Field Theories (NCFT) arise naturally. For reviews on early studies, see e.g \cite{dnsw-rev} and references therein. Compared to the ordinary field theories, NCFT have their own salient features. In particular, many efforts have been focused on the exploration of their quantum behavior in order to obtain a good understanding of their renormalisation properties. The renormalisation of NCFT is known to be often a difficult task since most of these theories are non-local, thus precluding the use of the standard machinery controlling the ordinary local field theories. The technical hard points may even be complicated by the possible appearance of the UV/IR mixing, a typical phenomenon of NCFT which spoils renormalisability. For the popular Moyal spaces $\mathbb{R}^4_\theta$ and $\mathbb{R}^2_\theta$ as well as for $\mathbb{R}^3_\lambda$, a deformation of $\mathbb{R}^3$ \cite{lagraa}, it has been shown that this phenomenon and all the technical difficulties can be overcome from different ways within some NCFT as well as some non-commutative gauge models leading to renormalisable (or even finite in some instance) field theories on these quantum spaces \cite{Grosse:2003aj-pc,brol-1,vign-sym,thes-vt,wal-16,poulwal-1,vitwal2,vitwal1} and outlining the deep relationship between NCFT and matrix models \cite{Wallet:2007c,matrix1,matrix2}. Recall that the 4-d Moyal space can be viewed {\it{informally}} as $\mathbb{C}[x_\mu]/\mathcal{R}$, the quotient of the free algebra generated by 4 hermitean coordinates $(x_\mu)_{\mu=1,...,4}$ by the relation $\mathcal{R}$ defined by $[x_\mu,x_\nu]=i\theta_{\mu\nu}$ where $\theta_{\mu\nu}$ is a skew symmetric constant tensor. This deformation of $\mathbb{R}^4$ can be described as a (suitable) algebra of functions on $\mathbb{R}^4$ equipped with the popular Moyal product \cite{groen,Moyal} obtained from the Wigner-Weyl quantization scheme. For various presentations of the Moyal product, see e.g \cite{dnsw-rev}. The non-commutative $\mathbb{R}^3_\lambda$,  another example of space with Lie algebra type non-commutativity, as the $\kappa$-Minkowski space-time also is, can be viewed {\it{informally}} as related to the universal enveloping algebra of $\mathfrak{su}(2)$, $U(\mathfrak{su}(2))\simeq\mathbb{C}[x_i]/\mathcal{R}'$, where the relation $\mathcal{R}'$ is defined by $[x_i,x_j]=i\varepsilon_{ijk}x_k$. For various derivations of star products related to $\mathbb{R}^3_\lambda$, see e.g \cite{lagraa,poulwal-1} and references therein. Note that these non-commutative spaces share a common underlying structure, each one being related to a group algebra. This latter corresponds, in the Moyal case, to the algebra for the Heisenberg group, which actually underlies the Weyl quantization, as it will be recalled below. For the space $\mathbb{R}^3_\lambda$, it is the convolution algebra of $SU(2)$, which has been shown to play an essential role in originating the special properties of $\mathbb{R}^3_\lambda$ \cite{vitwal1,vitwal2,wal-16}. In the case of $\kappa$-Minkowski space-time, the relevant group algebra is the convolution algebra of the affine group as it will be shown below.\bigskip

An important question to address is the fate of the symmetries of a non-commutative space-time. This has triggered a lot of works using various approaches which basically depend if one insists on preserving (almost all) the classical symmetries or if one considers deformed ones. For instance in \cite{Doplich1} the attention was focused on preserving the classical (undeformed) Lorentz or Poincar\'e symmetries for the Moyal space, as well as in \cite{dabrow} for $\kappa$-Minkowski space. In this latter work, the authors ensures classical covariance of $\kappa$-Minkowski space starting from a generalised version of it introduced in \cite{luk3}, i.e. $[x_\mu,x_\nu]=i\kappa^{-1}(v_\mu x_\nu - v_\nu x_\mu)$. They show that, under some assumptions, deformed (quantum) symmetries are not the only viable and consistent solution for treating such models. Note however that the original $\kappa$-Minkowski space \eqref{kappa-alg-d} (which we consider in this paper) does not fit in that description and breaks the classical relativity principle. This leads us to the other approach widely studied in the literature, namely the extension of the usual notion of Lie algebra symmetries to the one of (deformed) Hopf algebra symmetries aiming to encode the new (canonical) symmetries for the quantum space-times. This point of view is motivated by the fact that, in the commutative case, the Minkowski space-time can be regarded as the homogeneous space the Poincar\'e symmetry group acts on transitively. Hence, a deformation of the former should (in principle) implies a deformation of the latter and vice versa. This idea underlies the original derivation of $\kappa$-Minkowski as the homogeneous space associated to $\kappa$-Poincar\'e \cite{majid-ruegg}. Another interesting exemple (to put in perspective with \cite{Doplich1}) is given in \cite{chaichian}, where it is shown that the symmetries for the Moyal space can be obtained through formal (Drinfeld) twist deformation of the Lorentz sector of the Poincar\'e algebra while translation remains undeformed. General discussions on the fate of the Poincar\'e symmetries within the context of non-commutative (or quantum) space-times can be found in \cite{GAC2002:2} and references therein.\bigskip
 
NCFT on $\kappa$-Minkowski space have received a lot of interest from a long time, see for instance \cite{ital-1,habsb-imp,kappa-star1,hrvat-1}, but amazingly their quantum properties are not so widely explored, compared to the present status of the above mentioned NCFT. Nevertheless, the UV/IR mixing within some scalar field theories on $\kappa$-Minkowski has been examined a long time ago in \cite{gross-whl} and found to possibly occur. The corresponding analysis was based on a star product for the $\kappa$-deformation derived in \cite{star-spunz} from a general relationship between the Kontsevich formula and the Baker-Campbell-Hausdorff (BCH) formula that can be conveniently used when the non-commutativity is of Lie algebra type \cite{ypa}. NCFT considered in \cite{gross-whl} was $\kappa$-Poincar\'e invariant, which is a physically reasonable requirement, keeping in mind the important role played by the Poincar\'e invariance in ordinary field theories together with the fact that $\kappa$-Poincar\'e algebra can be viewed as describing the quantum symmetries of the $\kappa$-Minkowski space-time. \bigskip

It turns out that a very convenient star product for $\kappa$-Minkowski space can be obtained from a mere adaptation of the initial Wigner-Weyl quantization scheme which gives rise to the popular Moyal product. This can be illustrated schematically as follows. Recall that one important feature of this scheme is the notion of ``twisted convolution" of two functions{\footnote{with $f,g\in L^1(\mathbb{R}^2)$.}} $f$ and $g$ on the phase space $\mathbb{R}^2$, that we denote by $f\bullet g$, whose explicit expression was first given by von Neumann \cite{jvn}. This product is defined by $W(f\bullet g)=W(f)W(g)$ where $W(f)$ is the Weyl operator given by $W(f)=\int d\xi_1d\xi_2\ e^{i(\xi_1P+\xi_2Q)}f(\xi_1,\xi_2)$ in which the unitary operator in the integrand can be viewed as an element of the unimodular Heisenberg group{\footnote{To see that, use e.g the Glauber formula to reproduce the usual composition law for elements of the Heisenberg group.}}, obtained by exponentiating the Heisenberg algebra, says $[P,Q]=i\theta$ where $\theta$ is central. From this follows directly the Moyal product defining the deformation of $\mathbb{R}^2$. It is defined by $f\star g=\mathcal{F}^{-1}(\mathcal{F}f\bullet\mathcal{F}g)$ where the Weyl quantization map is $Q(f)=W(\mathcal{F}f)$ and $\mathcal{F}f$ is the Fourier transform of $f$.\bigskip

The natural extension of the above scheme to the construction of a star product for $\kappa$-Minkowski can then be achieved by simply replacing the Heisenberg group by the non-unimodular affine group as explained below, while $W(f)$ will be replaced by a representation of the convolution algebra of the affine group. Doing this, one can take advantage of the machinery of the harmonic analysis on Lie groups and, in particular, measures involved in action functionals are provided by Haar measures. Note that such a viewpoint has also been intensively used in \cite{vitwal2,wal-16} for $\mathbb{R}^3_\lambda$, the relevant group being $SU(2)$ reflecting the $\mathfrak{su}(2)$ non-commutativity of the quantum space and has provided the relationship between $\mathbb{R}^3_\lambda$ and the convolution algebra of $SU(2)$, the determination of the natural measure in the action functionals and by the way clarified the origin of the matrix basis used in \cite{vitwal1}. In the case of $\kappa$-Minkowski space, we note that such a natural construction has already been used in \cite{DS,matas} to derive a star product for a 2-dimensional $\kappa$-Minkowski space and to characterize a related multiplier algebra \cite{DS}. As far as we know, this product was amazingly not further exploited in the study of NCFT on $\kappa$-Minkowski space, despite its relatively simple expression and the associated tools of group harmonic analysis which make him well adapted to the study of quantum field theories. \bigskip

The construction of this natural star product defining the $\kappa$-deformation of the 4-d Minkowski space, considered with Euclidean signature in the present work, is presented in the section \ref{section2}. We then study in the section \ref{section3} different classes of $\kappa$-Poincar\'e invariant (complex) scalar field theories on the 4-d $\kappa$-Minkowski space whose commutative limit coincides with the usual $\phi^4$ theory. The kinetic operators are chosen to be square of Dirac operators. Requiring $\kappa$-Poincar\'e invariance forces the (Lebesgue) integral involved  in the actions to be a twisted trace with respect to the star product. This therefore defines a Kubo-Martin-Schwinger (KMS) weight on the non-commutative (C*-)algebra modeling the $\kappa$-Minkowski space. The associated modular group and Tomita modular operator are characterized. This is presented in the subsection \ref{section30} where we also discuss the possibility for the above KMS weight together with the associated modular data related on the non-commutative algebra modeling the $\kappa$-Minkowski space to generate the appearance of KMS states on the algebra of observables related to a global (observer-independent) time. The mathematical material as well as technical computations are collected in the appendix \ref{apendixB}.\\
The one-loop contributions to the 2-point functions of each of these theories are computed and their UV and IR behaviors are analyzed. The corresponding material is given in the subsections \ref{section31} and \ref{subsection32}. We find that the twist automorphism related to the twisted trace splits the planar contributions to the 2-point function into different IR finite contributions whose UV behavior is controlled by the twist. These contributions are found to be at most  UV linearly diverging, some being UV finite. A part of scalar theories considered in this work cannot give rise to non-planar contributions to the 2-point function so that these theories are expected to be free of UV/IR mixing. Conversely, UV/IR mixing shows up in another class of theories for which we find that the non-planar contributions to the 2-point function, while finite at  non zero external momenta, becomes singular  at exceptional zero external momenta with polynomial singularity. These results are finally discussed in the section \ref{section4}.

\section{$\kappa$-Minkowski space as a group algebra.}\label{section2}
\subsection{Convolution algebras and $\kappa$-Minkowski spaces.}\label{section2-1}
A convenient presentation of the $\kappa$-Minkowski space can be achieved by exploiting standard objects of the framework of group algebras and (C*-)dynamical systems \cite{dana}. This approach, which has been used in \cite{DS,matas} is the one we mainly follow in this paper. This framework has also been used in recent studies on $\mathbb{R}^3_\lambda$ spaces \cite{vitwal2,wal-16} related to the convolution algebra of the compact $SU(2)$ Lie group. Here, the relevant group is (related to) the affine group of the real line in the 2-dimensional case, i.e. a semi direct product of the two abelian groups $\mathbb{R}$, which extends in the $(d+1)$-dimensional case to $\mathbb{R}\ltimes_{\phi}\mathbb{R}^d$. We now collect the suitable material for the ensuing analysis.\bigskip

First, recall that the $\kappa$-deformation of the Minkowski space can be {\it{informally}} viewed as related to the universal enveloping algebra of the Lie algebra $\mathfrak{g}$ defined by:
\begin{equation}
[x_0,x_i]=\frac{i}{\kappa}x_i,\ \ [x_i,x_j]=0,\ \ i,j=1,\cdots, d.\label{kappa-alg-d}
\end{equation}
Here, $\kappa$ is a real number ($\kappa>0$) and the coordinates $x_0,\ x_i$ are assumed to be self-adjoint operators acting on some suitable Hilbert space. It turns out that $\mathfrak{g}$ is solvable. This can be easily deduced from the so-called derived Lie algebra $[\mathfrak{g},\mathfrak{g}]$ which is readily seen to be nilpotent. This is equivalent to have solvable $\mathfrak{g}$. Hence the associated Lie group, hereafter denoted by $\mathcal{G}_{d+1}$, is solvable (see e.g Theorem 5.9 of \cite{ypa-2}).  We use this property below to characterize the relevant algebra modelling the non-commutative space. \\
Notice that any Lie group of the form $A\ltimes_{\phi}B$, $\phi:A\to \mathrm{Aut}(B)$, where $A$ and $B$ are Abelian connected Lie groups, is solvable and connected (and is simply connected whenever $A$ and $B$ are simply connected). This is the case for $\mathcal{G}_{d+1}=\mathbb{R}\ltimes_{\phi}\mathbb{R}^d$, relevant to describe the $(d+1)$-dimensional $\kappa$-Minkowski spaces. This group is not unimodular signaling the existence of distinct left and right-invariant Haar measures, denoted respectively by $d\mu$ and $d\nu$. They are related by the modular function of $\mathcal{G}_d$, a continuous group homomorphism $\Delta_{\mathcal{G}_{d+1}}:\mathcal{G}_{d+1}\to\mathbb{R}^+_{/0}$, by $d\nu(s)=\Delta_{\mathcal{G}_{d+1}}(s^{-1})d\mu(s)$ for any $s\in\mathcal{G}_{d+1}$.\bigskip

For the moment, we assume $d=1$, the extension to $d=3$ is straightforward and will be exploited below. $\mathcal{G}_2$ is known to be the orientation-preserving affine group of the real line, i.e. the ``$(ax+b)$-group", $a>0$, widely studied in the mathematical literature. For basic mathematical 
details, see e.g \cite{dana,khalil} and references therein. For our present purpose, this (non-abelian simply connected) Lie group can be conveniently characterized by defining
\begin{equation}
W(p^0,p^1):=e^{ip^1x_1}e^{ip^0x_0}\label{group-elem},
\end{equation} 
where $p^0,\ p^1\in\mathbb{R}$ can be interpreted as momenta. The group elements \eqref{group-elem} are related to the more traditional exponential form of the Lie algebra \eqref{kappa-alg-d} through a mere redefinition of $p^1$. Indeed, by using in \eqref{group-elem} the simplified BCH formula $e^Xe^Y=e^{\lambda(u)X+Y}$, valid whenever $[X,Y]=uX$, see \cite{vanbrunt}, where $\lambda(u)=\frac{ue^u}{e^u-1}$, one obtains $W(p^0,p^1)=e^{i(p^0x_0+\lambda(\frac{p^0}{\kappa})p^1x_1)}$. However, \eqref{group-elem} is easier to manipulate for the ensuing computations. Now, upon using $e^Xe^Y=e^{Y}e^{e^uX}$ which holds true when again $[X,Y]=uX$, one obtains from \eqref{group-elem} the group product on $\mathcal{G}_2$ given by
\begin{equation}
W(p^0,p^1)W(q^0,q^1)=W(p^0+q^0,p^1+e^{-p^0/\kappa}q^1).\label{grouplaw}
\end{equation}
The unit element and inverse are respectively given by
\begin{equation}
\bbone_{G}=W(0,0),\ W^{-1}(p^0,p^1)=W(-p^0,-e^{p^0/\kappa}p^1)\label{group-unit-invers}.
\end{equation}
At this point, some remarks are in order. 
\begin{itemize}
\item {First, observe that the usual composition law for the $(ax+b)$-group can be obtained from \eqref{grouplaw} by representing the group elements \eqref{group-elem} as
\begin{equation}
W(p^0,b)=\begin{pmatrix}e^{-p^0/\kappa}&b\\0&1\end{pmatrix}
\end{equation}
and setting $a:=e^{-p^0/\kappa}$. That latter rewriting exhibits clearly the semi-direct product structure of $\mathcal{G}_2$ as
\begin{equation}
\mathcal{G}_2=\mathbb{R}^+_{/0}\ltimes_{\check{\phi}}\mathbb{R},\label{semi-dir}
\end{equation}
with $\check{\phi}:\mathbb{R}^+_{/0}\to\mathrm{Aut}(\mathbb{R})$ being given by the adjoint action of $\mathbb{R}^+_{/0}$ on $\mathbb{R}$. Indeed, the identifications $a\mapsto (a,0)$ and $b\mapsto (1,b)$ yield respectively the factors $\mathbb{R}^+_{/0}$ and $\mathbb{R}$ appearing in \eqref{semi-dir} while the action $\check{\phi}$, defined by $\check{\phi}(a)b=(a,0)(1,b)(a^{-1},0)$, is reflected at the level of \eqref{grouplaw} in
\begin{equation}
\phi:\mathbb{R}\to \mathrm{Aut}(\mathbb{R}),\qquad \phi(p^0)q=e^{-p^0/\kappa}q . \label{autom}
\end{equation}%
}
\item {Next, note that the energy-momentum composition law is essentially given by the BCH formula for the Lie group underlying the non-commutative space-times whose algebras of coordinates are of Lie algebra type. This is the case for $\kappa$-Minkowski, see eqn. \eqref{kappa-alg-d}, as well as for the Moyal plane (resp. $\mathbb{R}^3_\lambda$) whose algebra of coordinate operators is given by the Heisenberg algebra $[x_\mu,x_\nu]=i\theta$ (resp. $\mathfrak{su}(2)$ algebra $[x_\mu,x_\nu]=i\lambda\varepsilon_{\mu\nu}^{\hspace{9pt}\rho} x_\rho$). Here, the composition law can be directly read from \eqref{grouplaw} and reflects the non trivial coproduct structure of the $\kappa$-Poincar\'e algebra, see \eqref{hopf1}. 
}
\end{itemize}

Let $\pi_U:\mathcal{G}_2\to\mathcal{B}(\mathcal{H})$ denote a (strongly continuous) unitary representation of $\mathcal{G}_2$ where $\mathcal{H}$ is some suitable Hilbert space and $\mathcal{B}(\mathcal{H})$ is the (C*-)algebra of bounded operators on $\mathcal{H}$. A star product defining the 2-dimensional $\kappa$-Minkowski space can be obtained in a way similar to the usual Weyl quantization leading to the construction of Moyal product on the Moyal plane $\mathbb{R}^2_\theta$, see \cite{hennings}, the Heisenberg algebra and Heisenberg group being replaced now by \eqref{kappa-alg-d} and $\mathcal{G}_2$ \eqref{semi-dir} respectively. Accordingly, it is convenient to start from $L^1(\mathcal{G}_2)$, the convolution algebra of $\mathcal{G}_2$. Recall that it is a $^*$-algebra made of the set of integrable complex-valued functions on $\mathcal{G}_2$ with respect to some Haar measure equipped with the related convolution product{\footnote{Recall that $L^1(\mathcal{G}_2)$ is isomorphic to the completion w.r.t. the norm $||f||_1=\int_{\mathcal{G}_2}d\nu(s)f(s)$ of the algebra of compactly supported complex-valued functions on $\mathcal{G}_2$.}}. From now on, it will be assumed to be the right-invariant measure. Accordingly, the convolution product is defined by $(f\circ g)(t)=\int_{\mathcal{G}_2}d\nu(s)\ f(ts^{-1})g(s)$ for any $t\in\mathcal{G}_2$, $f, g\in{L^1(\mathcal{G}_2)}$. The involutive structure of the algebra can be ensured by any element of the one-parameter family of involutions defined $\forall t\in\mathcal{G}_2$ by $f^*(t):=\bar{f}(t^{-1})\Delta_{\mathcal{G}_2}^\alpha(t)$, $\alpha\in\mathbb{R}$. It turns out that the choice $\alpha=1$, assumed from now on, ensures that any representation of the convolution algebra defined for any $f\in{L^1(\mathcal{G}_2)}$ by
\begin{equation}
\pi:L^1(\mathcal{G}_2)\to \mathcal{B}(\mathcal{H}),\quad \pi(f)=\int_{\mathcal{G}_2}d\nu(s)f(s)\pi_U(s),\label{unireps}
\end{equation}
is a non-degenerate $^*$-representation. Indeed, a simple computation yields 
\begin{equation}
\langle u,\pi(f)^\dag v\rangle=\langle \pi(f)u,v\rangle=\int_{\mathcal{G}_2}d\nu(s)\bar{f}(s)\langle u,\pi_U(s^{-1})v \rangle,\label{equai} 
\end{equation}
where anti-linearity of the Hilbert product $\langle\cdot,\cdot\rangle$ and unitary property of $\pi_U$ have been used. Note that in \eqref{equai} the symbol $^\dag$ denotes the adjoint operation acting on operators, the nature of the various involutions should be obvious from the context. On the other hand, one computes
\begin{equation}
\langle u,\pi(f^*) v\rangle=\int_{\mathcal{G}_2}d\nu(s)\Delta_{\mathcal{G}_2}^\alpha(s)\bar{f}(s^{-1})\langle u,\pi_U(s)v\rangle,\label{equaii} 
\end{equation}
which combined with the relation $d\nu(s^{-1})=\Delta_{\mathcal{G}_2}(s)d\nu(s)$ is equal to \eqref{equai} provided $\alpha=1$.\bigskip

To summarize:
\begin{equation}
\pi(f)^\dag=\pi(f^*),
\end{equation}
and one can easily check that 
\begin{equation}
\pi(f\circ g)=\pi(f)\pi(g), \label{alg-morph-conv}
\end{equation}
for any $f, g\in{L^1(\mathcal{G}_2)}$.
\subsection{Quantization map and star product.}\label{2-2}
Let $\mathcal{F}f(p^0,p^1):=\int_{\mathbb{R}^2}dx_0dx_1e^{-i(p^0x_0+p^1x_1)}f(x_0,x_1)$ be the Fourier transform of $f\in L^1(\mathbb{R}^2)$. In the following, $\mathcal{S}_c$ denotes the space of Schwartz functions on $\mathbb{R}^2$ with compact support in the first variable.\bigskip

The quantization map is defined \cite{DS,matas} upon identifying functions on $\mathcal{G}_2$ with functions on $\mathbb{R}^2$ in view of \eqref{group-elem}-\eqref{group-unit-invers}. Namely, for any $f\in L^1(\mathbb{R}^2)\cap\mathcal{F}^{-1}(L^1(\mathbb{R}^2))$, we define
\begin{equation}
Q(f):=\pi(\mathcal{F}f),\label{quantizmap}
\end{equation}
where $\pi$ is the representation given by \eqref{unireps}. Notice that in view of \eqref{group-elem}, functions involved in the convolution product and involution map defined above are interpreted as Fourier transforms of functions of space-time coordinates. Hence, the occurrence of $\mathcal{F}f$ in the RHS of \eqref{quantizmap}. Then, since $Q$ must be a morphism of algebra, one writes 
\begin{equation}
Q(f\star g)=Q(f)Q(g)=\pi(\mathcal{F}f)\pi(\mathcal{F}g)=\pi(\mathcal{F}f\circ\mathcal{F}g)\label{intermed1}
\end{equation}
where \eqref{alg-morph-conv} has been used to obtain the last equality in \eqref{intermed1}, which compared with $Q(f\star g)=\pi(\mathcal{F}(f\star g))$ stemming from \eqref{quantizmap} and using the non-degeneracy of \eqref{unireps}, yields 
\begin{equation}
f\star g=\mathcal{F}^{-1}(\mathcal{F}f\circ\mathcal{F}g)\label{star prod-gene},
\end{equation}
where $\mathcal{F}^{-1}$ is the inverse Fourier transform on $\mathbb{R}^2$. In the same way, the requirement for $Q$ to be a $^*$-morphism yields 
\begin{equation}
f^\dag=\mathcal{F}^{-1}(\mathcal{F}(f)^*)\label{involgen}.
\end{equation}
Note that both the star product and the involution are representation independent despite the fact that the quantization map $Q$ depends on $\pi$.

Finally, by using the fact that the right-invariant measure on $\mathcal{G}_2$ is $d\nu(p^0,p^1)=dp^0dp^1$, i.e. the Lebesgue measure, with the modular function given by
\begin{equation}
\Delta_{\mathcal{G}_2}(p^0,p^1)=e^{p^0/\kappa},\label{modul-2d} 
\end{equation}
and combining the definition of the right-convolution product given above with eqns. \eqref{group-elem}-\eqref{group-unit-invers}, a simple calculation yields, for any $f,g\in\mathcal{F}(\mathcal{S}_c)$,
\begin{equation}
(f\star g)(x_0,x_1)=\int \frac{dp^0}{2\pi} dy_0\ e^{-iy_0p^0}f(x_0+y_0,x_1)g(x_0,e^{-p^0/\kappa}x_1)  \label{starpro},
\end{equation}
with $f\star g\in\mathcal{F}(\mathcal{S}_c)$, 
and 
\begin{equation}
f^\dag(x_0,x_1)= \int \frac{dp^0}{2\pi} dy_0\ e^{-iy_0p^0}{\bar{f}}(x_0+y_0,e^{-p^0/\kappa}x_1),\ \ f^\dag\in\mathcal{F}(\mathcal{S}_c)\label{invol},
\end{equation}
which coincide with the star product and involution of \cite{DS,matas}. \bigskip

\noindent At this point, some comments are in order. 
\begin{itemize}
\item First, it is instructive to get more insight on $C^*(\mathcal{G}_2)$, the C$^*$-algebra which models the $\kappa$-Minkowski space. Indeed, the completion of $L^1(\mathcal{G}_2)$ with respect to the norm related to the left regular representation on $L^2(\mathcal{G}_2)$ yields the reduced group C$^*$-algebra, $C^*_{red}(\mathcal{G}_2)$. Furthermore, since $\mathcal{G}_2$ is amenable as any solvable (locally compact) group, one has $C^*_{red}(\mathcal{G}_2)\simeq C^*(\mathcal{G}_2)$, involving as dense $^*$-subalgebra the set of Schwartz functions with compact support equipped with the above convolution product.
\item Eqns. \eqref{starpro} and \eqref{invol} can be extended \cite{DS} to (a subalgebra of) the multiplier algebra{\footnote{It involves the smooth functions on $\mathbb{R}^2$ satisfying standard polynomial bounds together with all the derivatives, with Fourier transform having compact support in the first variable.}} of $\mathcal{F}(\mathcal{S}_c)$ involving in particular $x_0$ and $x_1$ and the unit function. From \eqref{starpro} and \eqref{invol}, one easily obtains 
\begin{equation}
x_0\star x_1=x_0x_1+\frac{i}{\kappa}x_1,\ x_1\star x_0=x_0x_1,\ x_\mu^\dag=x_\mu,\ \mu=1,2\label{def-relations},
\end{equation}
consistent with the defining 
relation \eqref{kappa-alg-d} (for $d=1$).
\end{itemize}
The extension of the above construction to the 4-dimensional case is straightforward. Indeed, the group law becomes now $W(p^0,\vec{p})W(q^0,\vec{q})=W(p^0+q^0,\vec{p}+e^{-p^0/\kappa}\vec{q})$ with $W(p^0,\vec{p}):=e^{ip^ix_i}e^{ip^0x_0}$, $\vec{p}=(p^i,\ i=1,2,3)$ and $W^{-1}(p^0,\vec{p})=W(-p^0,-e^{p^0/\kappa}\vec{p})$. This entails the semi-direct product structure $\mathcal{G}_4=\mathbb{R}\ltimes_{\phi}\mathbb{R}^3$ where $\phi$ is still given by \eqref{autom}. Then, the construction leading to \eqref{starpro} and \eqref{invol} can be thoroughly reproduced, replacing $\mathbb{R}^2$ by $\mathbb{R}^4$ and \eqref{modul-2d} by
\begin{equation}
\Delta_{\mathcal{G}_4}(p^0,\vec{p})=e^{3p^0/\kappa}\label{modul-4d}.
\end{equation}
Setting for short $x:=(x_0,\vec{x})$, one obtains
\begin{align}
(f\star g)(x)&=\int \frac{dp^0}{2\pi} dy_0\ e^{-iy_0p^0}f(x_0+y_0,\vec{x})g(x_0,e^{-p^0/\kappa}\vec{x})  \label{starpro-4d},\\
f^\dag(x)&= \int \frac{dp^0}{2\pi} dy_0\ e^{-iy_0p^0}{\bar{f}}(x_0+y_0,e^{-p^0/\kappa}\vec{x})\label{invol-4d},
\end{align}
for any functions $f,g\in\mathcal{F}(\mathcal{S}_c)$ and one still has $f\star g\in\mathcal{F}(\mathcal{S}_c)$ and $f^\dag\in\mathcal{F}(\mathcal{S}_c)$. Here, $\mathcal{S}_c$ is now the set of Schwartz functions of $\mathbb{R}^4$ with compact support in the $p^0$ variable. Of course, comments similar to the one given above for $C^*(\mathcal{G}_2)$ and \eqref{def-relations} apply to the 4-dimensional case on which we focus in the rest of this paper. Notice that the functions in $\mathcal{F}(\mathcal{S}_c)$ are by construction analytic in the variable $x_0$, being Fourier transforms of functions with compact support in the variable $p^0$, thanks to the Paley-Wiener theorem.\bigskip

For the ensuing discussion, it will be sufficient to consider the algebra $\mathcal{F}(\mathcal{S}_c)$ unless otherwise stated, which will be denoted hereafter by $\mathcal{M}_\kappa$.
\subsection{$\kappa$-Poincar\'e invariant actions.}\label{section22}
In this section, we discuss general properties shared by $\kappa$-Poincar\'e invariant action functionals for complex-valued scalar fields, denoted generically by $S_\kappa(\phi)$.\\
Let $\mathcal{P}_\kappa$ denote the $\kappa$-Poincar\'e algebra. We will demand that the action functional $S_\kappa(\phi)$ obeys the following two conditions:
\begin{enumerate}
\item $S_\kappa(\phi)$ is $\mathcal{P}_\kappa$-invariant which is expressed as
\begin{equation}
h\triangleright S_\kappa(\phi)=\epsilon(h)S_\kappa(\phi),\label{invarquant}
\end{equation}
for any $h$ in the Hopf algebra $\mathcal{P}_\kappa$ where $\epsilon$ is the co-unit of $\mathcal{P}_\kappa$ (see appendix \ref{apendixA}), 
\item $S_\kappa(\phi)$ reduces to the standard $\phi^4$ scalar field theory in the commutative limit $\kappa\to\infty$.
\end{enumerate}
Recall that $\mathcal{P}_\kappa$ has a natural action on $\mathcal{M}_\kappa$ which informally may be viewed as the action of a quantum symmetry on the corresponding quantum (non-commutative) space modelled by $\mathcal{M}_\kappa$, reflecting the fact that the algebra $\mathcal{M}_\kappa$ is a left-module over the Hopf algebra $\mathcal{P}_\kappa$. A convenient presentation of $\mathcal{P}_\kappa$ can be obtained from the 11 elements $(P_i, N_i,M_i, \mathcal{E},\mathcal{E}^{-1})$, $i=1,2,3$, which are respectively the momenta, the boost and the rotations together with the invertible element
\begin{equation}
\mathcal{E}:=e^{-P_0/\kappa},\label{Erond} 
\end{equation}
to be discussed at length in a while. The relations between these elements which characterize the Hopf algebra structure together with the duality between the Hopf subalgebra describing the ``deformed translation algebra" and the $\kappa$-Minkowski space are presented in the appendix \ref{apendixA} for the sake of completeness.\bigskip

Going back to the condition a), it is known that the invariance of $S_\kappa(\phi)$ under $\mathcal{P}_\kappa$ is automatically achieved by considering action functionals of the form 
\begin{equation}
S_\kappa(\phi)=\int d^4x\ \mathcal{L}(\phi),\label{weight}
\end{equation}
where $\phi\in\mathcal{F}(\mathcal{S}_c)$ so that $\mathcal{L}(\phi)\in\mathcal{F}(\mathcal{S}_c)$ in view of \eqref{starpro}. Indeed, by using \eqref{left-module0}-\eqref{left-modules1bis}, one has plainly 
\begin{equation}
P_\mu\triangleright S_\kappa(\phi):=\int d^4x\ P_\mu \triangleright \mathcal{L}(\phi)=0,\ \ M_i\triangleright S_\kappa(\phi):=\int d^4x\ M_i \triangleright \mathcal{L}(\phi)=0,
\end{equation}
while 
\begin{equation}
\mathcal{E}\triangleright S_\kappa(\phi):=\int d^4x\ \mathcal{E}\triangleright \mathcal{L}(\phi)=S_\kappa(\phi)
\end{equation}
where the last equality stems from the Cauchy theorem. Next, one obtains from \eqref{left-module2} 
\begin{equation}
N\triangleright S_\kappa(\phi):=\int d^4x\ ([\frac{\kappa}{2}L_{x_i}(\mathcal{E}-\mathcal{E}^{-1})+L_{x_0}P_i\mathcal{E}+L_{x_i}\vec{P}^2\mathcal{E}])\triangleright \mathcal{L}(\phi).
\end{equation}
By using \eqref{left-module0}, \eqref{left-module1}, one easily checks that the last two terms in the right hand side vanish as integrals of total derivative of Schwartz functions while the 2 contributions of the first term balance each other thanks to the Cauchy theorem.\bigskip

For further use, we quote useful formulas
\begin{align}
\int d^4x\ (f\star g^\dag)(x)&=\int d^4x\ f(x){\bar{g}}(x),\label{algeb-1}\\
\int d^4x\ f^\dag(x)&=\int d^4x\ {\bar{f}}(x),\label{int-form1}
\end{align}
stemming from mere changes of variables and the use of the Cauchy theorem as it can be easily verified. We note that a mere consequence of \eqref{algeb-1} is
\begin{equation}
\int d^4x\ f\star f^\dag\ge0,\ \ \int d^4x\ f^\dag\star f\ge0,\label{positivity}
\end{equation}
thus defining two positive maps $\int d^4x:\mathcal{M}_{\kappa+}\to \mathbb{R}^+$ where $\mathcal{M}_{\kappa+}$ denotes the set of positive elements of $\mathcal{M}_\kappa$.\bigskip

It turns out that the Lebesgue integral does not define a trace. Indeed, a simple computation yields
\begin{equation}
\int d^4x\ f\star g=\int d^4x\ (\sigma\triangleright g)\star f\label{twistrace},
\end{equation}
where we define for further convenience
\begin{equation}
\sigma\triangleright f:=\mathcal{E}^3\triangleright f=e^{-\frac{3P_0}{\kappa}}\triangleright f\label{twistoperator},
\end{equation}
in which $\mathcal{E}$ is given by \eqref{Erond}. Hence, the Lebesgue integral  cannot define a trace since cyclicity is lost in view of \eqref{twistrace}. Instead, it defines a twisted trace. \bigskip

Recall that a twisted trace (on an algebra) is defined in the mathematical literature as a linear positive map $\textrm{Tr}$ satisfying $\textrm{Tr}(a\star b)=\textrm{Tr}((\sigma\triangleright b)\star a)$, where $\sigma$ is an automorphism of the algebra called the twist. This is verified by the Lebesgue integral in view of \eqref{positivity}, \eqref{twistrace} where the corresponding twist is explicitly given by \eqref{twistoperator}, which will be discussed in the subsection \ref{section30}. This loss of cyclicity has often been considered as a troublesome feature of $\kappa$-Poincar\'e invariant field theories, this having probably discouraged the pursue of many studies of their properties at the quantum level.\bigskip

However, whenever there is a twisted trace, there is a related KMS condition (up to additional technical requirements that will not be essential for the ensuing discussion), a fact that is known in the mathematical literature. The relevant technical material needed for the discussion is presented in the appendix \ref{apendixB}. In the subsection \ref{section30},  we discuss some possible consequences of this KMS condition on field theories on $\kappa$-Minkowski space. In the subsections \ref{section31} and \ref{subsection32}, we construct a family of scalar field theories on 4-d $\kappa$-Minkowski space and study the UV and IR behaviour of the corresponding 2-point functions at one-loop order.

\section{Scalar field theories on 4-d $\kappa$-Minkowski space.}\label{section3}
\subsection{Trading cyclicity for KMS condition.}\label{section30}
To see that the Lebesgue integral actually defines a twisted trace, one key observation is to notice that \eqref{twistrace} and \eqref{twistoperator} can be interpreted as a KMS weight on $\mathcal{M}_\kappa$ for the group of $^*$-automorphisms of $\mathcal{M}_\kappa$
defined by
\begin{equation}
\sigma_t(f):=e^{it\frac{3P_0}{\kappa}}\triangleright f=e^{\frac{3t}{\kappa}\partial_0}\triangleright f, \label{sigmat-modul}
\end{equation}
for any $t\in\mathbb{R}$ and $f\in\mathcal{M}_\kappa$. This group of automorphisms is called the modular group for the KMS weight{\footnote{Roughtly speaking, a weight differs only from a state by an overall normalisation.}}. The corresponding mathematical details, technical computations and related references are collected in the appendix \ref{apendixB}. \\
The modular group, whose \eqref{sigmat-modul} is an example, is at the corner stone of the Modular Theory of Tomita-Takesaki, an essential tool in the area of von Neumann algebras. For details, see e.g \cite{takesaki} and references therein. It turns out that one of the initial motivations of Tomita to construct the Modular Theory was related to the harmonic analysis of (locally compact) non-unimodular group, as the one underlying the present study. In particular for these groups, the word ``modular'' refers to the modular function of the group, here \eqref{modul-4d}, while the Tomita modular operator  is simply the multiplication by the modular function \eqref{modul-4d}. Recall that Modular Theory, KMS condition and twisted trace are rigidly linked. Hence, it is not surprising that these structures underlie the present framework since the requirement of $\kappa$-Poincar\'e invariance of the action functional fixes the trace involved in it to be twisted. \bigskip

To summarize the analysis of appendix \ref{apendixB}, the KMS weight $\varphi$ is simply given by the map 
\begin{equation}
\varphi(f):=\int d^4x\ f(x),\label{varphi}
\end{equation}
for any $f\in\mathcal{M}_\kappa$ which verifies
\begin{equation}
\varphi(\sigma_t f)=\varphi(f),\ \ \varphi\big((e^{i\frac{3}{2\kappa}\partial_0}\triangleright f)\star(e^{-i\frac{3}{2\kappa}\partial_0}\triangleright f^\dag)\big)=\varphi(f^\dag\star f).\label{defining-properties}
\end{equation}
These 2 properties are actually defining properties for a KMS weight. Note that the $\kappa$-Poincar\'e invariance is crucial to insure that \eqref{defining-properties} holds true. It follows obviously that any action functional for a $\kappa$-Poincar\'e invariant theory is related to a KMS weight. Hence, the requirement of $\kappa$-Poincar\'e invariance trades the cyclicity of the trace for a KMS condition.\bigskip

Now, from a general theorem (Theorem [6.36] of 1st of ref. \cite{kuster}), one concludes that $\varphi$ must obey a KMS condition. Indeed, one defines{\footnote{ Note that $f_{a,b}(t)$ is continuous and bounded owing to the properties of the star product \eqref{starpro-4d}. As already mentioned at the end of Section \ref{2-2}, analyticity of $f_{a,b}$ stems from the Paley-Wiener theorem.}}
\begin{equation}
f_{a,b}(t):=\int d^4x\ \sigma_t(a)\star b,\label{eq0}
\end{equation}
for any $a,b\in\mathcal{M}_\kappa$. Then, by using the algebraic properties of the twist and $\sigma_t$, one computes
\begin{equation}
f_{a,b}(t)=\int d^4x\ \sigma_t(a)\star b =\int d^4x\ \sigma_i(b)\star \sigma_t(a)
=\int d^4x\ \sigma_i(b\star \sigma_{t-i}(a))=\int d^4x\ b\star \sigma_{t-i}(a).\label{eq1}
\end{equation}
in which we used \eqref{b10}. From this follows that
\begin{equation}
f_{a,b}(t+i)=\int d^4x\ b\star \sigma_{t}(a),\label{eq2}
\end{equation}
which verifies the above mentioned theorem (see appendix \ref{apendixB}). \bigskip

As pointed out in the appendix \ref{apendixB}, \eqref{eq0} and \eqref{eq2} represent an abstract version of the KMS condition introduced a long time ago as a tool to characterize equilibrium temperature states of quantum systems in field theory and statistical physics. To see that, set formally $f_{a,b}(t)=\langle \sigma_t(a)\star b\rangle$; then, \eqref{eq1} implies $\langle \sigma_t(a)\star b\rangle=\langle b\star\sigma_{t-i}(a)\rangle$ which bears some formal similarity with the usual form of the KMS condition for the quantum systems. Notice that $\sigma_t$ actually represents a ``time-translation operator" related to the Tomita operator $\Delta_T=e^{3P_0/\kappa}$ via $\sigma_t=(\Delta_T)^{it}$, as shown at the end of the appendix \ref{apendixB}. \bigskip

However, in the case of quantum systems or quantum field theory, $f_{A,B}(t)$ corresponds to a correlation function $\langle \Sigma_t(A)B\rangle_{\Omega}$ computed for some (thermal) vacuum $\Omega$ where $A$ and $B$ are now function(al)s (operators) of the fields and $\Sigma_t$ is the (Heisenberg) evolution operator, hence elements pertaining to the algebra of observables of the theory. But whenever a KMS condition holds true on the {\it{algebra of observables of a quantum system or a quantum field theory}}, the flow generated by the modular group, i.e. the Tomita flow, may be used to define a global (observer-independent) time which can be interpreted as the ``physical time". This reflects the deep correspondence between KMS condition and dynamics. This observation underlies the interesting proposal about the thermal origin of time introduced in \cite{ConRove}.\bigskip

While it  would be tempting to interpret $\sigma_t$ \eqref{sigmat-modul} as defining (or generating) a ``physical time" for the present system, akin to the thermal time mentioned above, no conclusion can yet be drawn. In fact, eqn. \eqref{KMS-abst} linked to the modular group and its associated KMS condition \eqref{eq1}, \eqref{eq2} only holds at the level of $\mathcal{M}_\kappa$, the algebra modelling the $\kappa$-Minkowski space. To show that a natural global time can be defined requires to determine if \eqref{eq1}, \eqref{eq2} force a KMS condition to hold true at the level of the algebra of observables. This could be achieved by actually showing the existence of some KMS state(s) on this latter algebra built from the path integral machinery. In view of the possibility to associate to $\kappa$-Poincar\'e invariant non-commutative field theories a natural global time, a physically appealing property, the implications of the KMS condition \eqref{eq1}, \eqref{eq2} shared by all these theories obviously deserves further study. The full analysis is beyond the scope of the present paper. \bigskip

We now pass to the construction of reasonable $\kappa$-Poincar\'e invariant action functionals and the study of the UV and IR property of their one-loop 2-point functions, adopting the standard viewpoint of the non-commutative field theories, namely representing the non-commutative action functional as an action functional describing non-local commutative field theories. It turns out that the use of the star product introduced in the section \ref{section2} simplifies the computations of the correlation functions. This will be exemplified by explicit computations of 2-point functions in the subsection \ref{subsection32}. We first introduce the main elements of our framework and analyse carefully the corresponding properties.

\subsection{Construction of real action functionals.}\label{section31}
\subsubsection{Preliminary considerations.}\label{section311}
It is convenient to begin by introducing the following Hilbert product on $\mathcal{M}_\kappa$
\begin{equation}
\langle f,g\rangle:=\int d^4x\left(f^\dag\star g\right)(x)=\int d^4x\ {\bar{f}}(x)(\sigma\triangleright g)(x),\  \forall f,g\in\mathcal{M}_\kappa. \label{hilbert-product}
\end{equation}
To check that \eqref{hilbert-product} defines actually a Hilbert product, one observes that positivity is apparent from \eqref{positivity} while ${\overline{\langle f,g \rangle}}=\langle g,f \rangle $ stems from \eqref{int-form1} applied to $f^\dag\star g=(g^\dag\star f)^\dag$. Furthermore, the corresponding Hilbert space $\mathcal{H}$ can be shown to be (unitarily) isomorphic to $L^2(\mathbb{R}^4)$, i.e. $\mathcal{H}\simeq L^2(\mathbb{R}^4)$. The proof is given in the appendix \ref{apendixC}.\\
One can verify that $P_i$, $i=1,2,3$, and $\mathcal{E}$ are self-adjoint w.r.t. the Hilbert product \eqref {hilbert-product}. Indeed, one computes
\begin{align}
\langle f,P_i^\dag\triangleright g\rangle&=\langle P_i\triangleright f,g \rangle=-\int d^4x\ (\mathcal{E}^{-1}P_i\triangleright(f^\dag))\star g=-\int d^4x\  (P_i\triangleright(f^\dag))\star (\mathcal{E}\triangleright g)\nonumber\\
&=\int d^4x\  (\mathcal{E}\triangleright(f^\dag))\star (P_i\mathcal{E}\triangleright g)=\int d^4x\  f^\dag\star (P_i\triangleright g)\nonumber\\
&=\langle f,P_i\triangleright g\rangle,
\end{align}
where we have successively used \eqref{dag-hopfoperat}, the $\kappa$-Poincar\'e invariance \eqref{invarquant}, \eqref{deriv-twist1} and \eqref{relation-calE}. Hence $P_i$ is self-adjoint. Self-adjointness of $P_0$ and $\mathcal{E}$ can be shown similarly.\bigskip

In order to construct real action functionals, notice that \eqref{hilbert-product} is $\mathbb{R}$-valued for any $f, g\in\mathcal{F}(\mathcal{S}_c)$ satisfying $\langle f,g\rangle=\langle g,f\rangle $. Hence, reality condition for kinetic term of the form $\langle f,K_\kappa f\rangle$ is automatically verified providing that the kinetic operator $K_\kappa$ (assumed in the following to have dense domain in $\mathcal{H}$) is self-adjoint since this implies $\langle f,K_\kappa f\rangle=\langle K_\kappa f,f\rangle $.\\
We further assume the kinetic operator $K_\kappa$ to be a pseudo-differential operator, i.e.
\begin{equation}
(K_\kappa f)(x)=\int \frac{d^4p}{(2\pi)^4} d^4y\ \mathcal{K}_\kappa(p)f(y)e^{ip(x-y)},\label{pseudoper}
\end{equation}
for any $f$ in the domain of $K_\kappa$, where $\mathcal{K}_\kappa(p)$ is some rational fraction of $(p^0,\vec{p})$. Note that self-adjointness for $K_\kappa$ requires
\begin{equation}
{\overline{\mathcal{K}_\kappa}}(p^0,\vec{p})=\mathcal{K}_\kappa(p^0,\vec{p})\label{constr-K},
\end{equation}
a condition that will be fulfilled in the situations we will consider below. Indeed, a simple computation yields
\begin{equation}
\langle f,K_\kappa f\rangle=\int d^4xd^4y\frac{d^4p}{(2\pi)^4}\ \bar{f}(x)f(y)e^{ip(x-y)}e^{-3p^0/\kappa}\mathcal{K}_\kappa(p^0,\vec{p}),\label{algeb2}
\end{equation}
while
\begin{equation}
\langle K_\kappa f,f\rangle=\int d^4xd^4y\frac{d^4p}{(2\pi)^4}\ \bar{f}(x)f(y)e^{ip(x-y)}e^{-3p^0/\kappa}{\overline{\mathcal{K}_\kappa}}(p^0,\vec{p}),\label{algeb3}
\end{equation}
proving the above statement.\bigskip

We are now in position to construct $\kappa$-Poincar\'e invariant action functionals $S_\kappa(\phi^\dag,\phi)$ such that 
\begin{equation}\label{action-limit}
\lim_{\kappa\to\infty}S_\kappa(\phi^\dag,\phi)=\int d^4x \left(\bar{\phi}(-\partial_\mu\partial^\mu+m^2)\phi+\lambda\bar{\phi}\phi\bar{\phi}\phi\right)(x),\ \lambda\in\mathbb{R},
\end{equation}
 i.e. fulfilling the condition b) introduced in the section \ref{section22}. We assume the following usual generic form for the action functionals 
\begin{equation}\label{star-act}
S_\kappa(\phi^\dag,\phi)=S_\kappa^\text{kin}(\phi^\dag,\phi)+S_\kappa^\text{int}(\phi^\dag,\phi),
\end{equation}
where $S_\kappa^\text{int}(\phi^\dag,\phi)$ is a quartic polynomial in the fields and $S_\kappa^\text{kin}(\phi^\dag,\phi)$ is the kinetic term. For the theories under consideration, the mass dimension for the fields and parameters are respectively $[\phi]=[\phi^\dag]=1$, $[\lambda]=0$ and $[m]=1$.
\subsubsection{Derivation of the kinetic term.}
Let us first discuss the kinetic term $S_\kappa^\text{kin}(\phi^\dag,\phi)$.\\According to the above discussion, admissible real kinetic terms are of the form 
\begin{equation}
\langle \phi, K_\kappa\phi \rangle,\ \ \langle \phi^\dag,K_\kappa\phi^\dag \rangle,\label{realkineticterm}
\end{equation}
where $K_\kappa$ is self-adjoint. Its explicit expression will be given in a while. We also incorporate all possible ``mass terms" of similar forms, namely $m^2\langle \phi,\phi \rangle$ and $m^2\langle\phi^\dag,\phi^\dag \rangle$.\bigskip

A first natural choice for the kinetic operator is provided by the first Casimir of the $\kappa$-Poincar\'e algebra $\mathcal{P}_\kappa$. This latter is given in the Majid-Ruegg basis by
\begin{equation} 
\mathcal{C}_\kappa(P_\mu)=4\kappa^2\sinh^2\left(\frac{P_0}{2\kappa}\right) + e^{P_0/\kappa} \vec{P}^2 \label{Casimir},
\end{equation}
or equivalently
\begin{equation}\label{Casimir2}
\mathcal{C}_\kappa(P_\mu)= e^{P_0/\kappa}\left(\kappa^2\left(1-e^{-P_0/\kappa}\right)^2+\vec{P}^2\right).
\end{equation}
The Casimir operator \eqref{Casimir} can be put into the form
\begin{equation}
\mathcal{C}_\kappa(P_\mu)=D_0^2+D_iD^i,\label{casimir-carre}
\end{equation}
with
\begin{equation}
D_0:=\kappa \mathcal{E}^{-1/2}(1-\mathcal{E}),\ \ D_i:= \mathcal{E}^{-1/2}P_i,\ i=1,2,3\label{dirac},
\end{equation}
where $D_0$ and $D_i$ define self-adjoint operators. To see that, first observe that one has $\int d^4x\ D_\mu f=0$ for any $f\in\mathcal{M}_\kappa$ and use \eqref{pairing-involution}, \eqref{invarquant} and \eqref{deriv-twist1} to compute for instance
\begin{align}
\langle D_if,g \rangle&=\int d^4x\ (D_if)^\dag\star g=-\int d^4x\ (\mathcal{E}^{-1/2}P_i\triangleright f^\dag)\star g\nonumber\\
&=-\int d^4x\ (P_i\triangleright f^\dag)\star
(\mathcal{E}^{1/2}\triangleright g)=\int d^4x\  f^\dag\star
(\mathcal{E}^{-1/2}P_i\triangleright g)\nonumber\\
&=\langle f,D_ig \rangle,
\end{align}
for any $f,g\in\mathcal{M}_\kappa$. The computation for $D_0$ is similar. Note by the way that $D_0$ and $D_i$ are not derivations of the algebra $\mathcal{M}_\kappa$. \bigskip

A second possible natural choice is given by the square of the equivariant Dirac operator involved in the construction of an equivariant spectral triple for the $\kappa$-Minkowski space \cite{frans-2}. It is given by
\begin{equation}
K^{eq}_\kappa(P_\mu):=\mathcal{C}_\kappa(P_\mu)+\frac{1}{4\kappa^2}\mathcal{C}_\kappa(P_\mu)^2.\label{kinetic}
\end{equation}
For latter convenience, we quote here a useful factorization of the kinetic operator \eqref{kinetic} supplemented by a mass term, assuming $0\leq m \leq \kappa$,
\begin{equation}
K^{eq}_\kappa(P_\mu)+m^2 = \frac{e^{2P_0/\kappa}}{4\kappa^2}\left(\vec{P}^{\hspace{2pt}2}+\kappa^2\mu^2_{+}\right)\left(\vec{P}^{\hspace{2pt}2}+\kappa^2\mu^2_{-}\right),\label{factor-propa}
\end{equation}
where the two positive functions $\mu^2_{+}$ and $\mu^2_{-}$ are given by
\begin{equation}
\mu^2_{\pm}(m,P_0):=\left[1 \pm 2e^{-P_0/\kappa} \sqrt{1-\left(\frac{m}{\kappa}\right)^2} + e^{-2P_0/\kappa}\right].
\end{equation}
Again, one can write
\begin{equation}
K^{eq}_\kappa(P_\mu)=D^{eq}_0D^{eq}_0+\sum_i D^{eq}_iD^{eq}_i,\label{carre-equiv}
\end{equation}
where
\begin{equation}
D^{eq}_0:=\frac{\mathcal{E}^{-1}}{2}\left(\kappa(1-\mathcal{E}^2)-\frac{1}{\kappa}\vec{P}^2\right) \ \ ,\ D^{eq}_i:=\mathcal{E}^{-1}P_i,\label{equivdirac}
\end{equation}
which can be easily verified to be self-adjoint using successively eqn. \eqref{pairing-involution}, the $\kappa$-Poincar\'e invariance \eqref{invarquant} and the twisted Leibnitz rules for the $P_\mu$ \eqref{deriv-twist1}, \eqref{deriv-twist2}.\bigskip

Now, recall that one has for any of the operators \eqref{dirac}, \eqref{equivdirac}, the following usefull formula 
\begin{equation}
\langle \mathcal{D}_\mu f,g\rangle=\langle f,\mathcal{D}_\mu g \rangle,\label{byparts}
\end{equation}
for any $f,g\in\mathcal{M}_\kappa$, in which $\mathcal{D}_\mu$ denotes any of the operators \eqref{dirac}, \eqref{equivdirac}, stemming from the self-adjointness of these operators. From \eqref{realkineticterm}, \eqref{casimir-carre}, \eqref{carre-equiv} and \eqref{byparts}, a suitable form for the kinetic term is then given by
\begin{align}
S^\text{kin}_\kappa(\phi^\dag,\phi)&=\langle \phi, (K_\kappa+m^2)\phi \rangle+\langle \phi^\dag,(K_\kappa+m^2)\phi^\dag \rangle\nonumber\\
&=\int d^4x\ \phi^\dag\star(K_\kappa+m^2)\phi+\int d^4x\ \phi\star(K_\kappa+m^2)\phi^\dag\nonumber\\
&=\int d^4x\ \phi^\dag\star(1+\sigma^{-1})(K_\kappa+m^2)\phi,\label{kinetic-term}
\end{align}
where $K_\kappa=\mathcal{D}_\mu\mathcal{D}^\mu$ is any of the 2nd order operators \eqref{Casimir}, \eqref{kinetic}. Note by the way that, ignoring the mass terms, one has 
\begin{equation}
\langle \phi, K_\kappa\phi \rangle=\langle \mathcal{D}_\mu\phi, \mathcal{D}_\mu\phi \rangle,
\end{equation}
and similarly for $\phi\to\phi^\dag$.\bigskip

It is important to realise that the analysis of the quantum behaviour of the NCFT under consideration can be conveniently carried out within the present framework by expressing the non-commutative action functional $S_\kappa(\phi^\dag,\phi)$ (involving star products) as a non-local ordinary quantum field theory $S_\kappa(\bar{\phi},\phi)$ (involving pointwise products). This can be achieved by making use of the integral forms for the star product \eqref{starpro-4d} and the involution \eqref{invol-4d} in the expression for the action functional \eqref{star-act}, \eqref{kinetic-term}, \eqref{interac-type1} and \eqref{interac-type2}.\\Applying this procedure to $S^\text{kin}_\kappa$ leads to a great simplification in the computation of the propagator, despite the fact that the star product \eqref{starpro-4d} is not closed w.r.t. the Lebesgue integral, i.e. $\int d^4x\ (f\star g)(x)\ne\int d^4x\ f(x)g(x)$. Indeed, further using \eqref{pseudoper}, we obtain 
\begin{align}
S^\text{kin}_\kappa(\bar{\phi},\phi) &= \int d^4x_1d^4x_2 \ \bar{\phi}(x_1)\phi(x_2) \mathcal{K}_\kappa(x_1-x_2),\label{pratik-kinet-act}\\
\text{with}\ \ \mathcal{K}_\kappa(x_1-x_2) &= \int \frac{d^4p}{(2\pi)^4} \left(1+e^{-3p^0/\kappa}\right) \left(\mathcal{K}_\kappa(p)+m^2\right) e^{ip\cdot(x_1-x_2)}.\label{pratik-kinet}
\end{align}
The corresponding propagator can be derived by solving $\int d^4yd^4z\ \mathcal{K}_\kappa(x-y)P_\kappa(y-z)f(z)=
\int d^4z\ \delta(x-z)f(z)$ for any suitable test function $f(z)$, which amounts to invert $\mathcal{K}_\kappa(x-y)$.\\
Finally, assuming $K_\kappa=K^{eq}_\kappa$, eqn. \eqref{kinetic}, one obtains
\begin{equation}
P_\kappa(x_1-x_2) = \int \frac{d^4p}{(2\pi)^4} \ \frac{e^{ip\cdot(x_1-x_2)}}{\left(1+e^{-3p^0/\kappa}\right) \left(\mathcal{K}^{eq}_\kappa(p)+m^2\right)},\label{intermed2}
\end{equation}
which, combining \eqref{intermed2} with \eqref{factor-propa}, yields
\begin{equation} \label{propagator}
P_{\kappa}(x_1-x_2)=\int \frac{d^4p}{(2\pi)^4} \ \frac{e^{-2p^0/\kappa}}{1+e^{-3p^0/\kappa}} \  \frac{(2\kappa)^2e^{ip\cdot(x_1-x_2)}}{\left(\vec{p}^{\hspace{2pt}2}+\kappa^2\mu^2_{+}\right)\left(\vec{p}^{\hspace{2pt}2}+\kappa^2\mu^2_{-}\right)}.
\end{equation}
While assuming $K_\kappa=\mathcal{C}_\kappa$, eqn. \eqref{Casimir}, leads in a similar manner to
\begin{align}
P_{\kappa}(x_1-x_2)&=\int \frac{d^4p}{(2\pi)^4} \ \frac{e^{-p^0/\kappa}}{1+e^{-3p^0/\kappa}}\ \frac{e^{ip\cdot(x_1-x_2)}}{\vec{p}^{\hspace{2pt}2}+\kappa^2 \mu^2},\\
\mu^2(m,p^0)&=\left(m/\kappa\right)^2 e^{-p^0/\kappa}+\left(1-e^{-p^0/\kappa}\right)^2.
\end{align}
\subsubsection{Derivation of the interaction term.}
Let us now discuss the interaction part $S^\text{int}_\kappa(\phi^\dag,\phi)$.\\
In view of \eqref{hilbert-product}, the requirement for the action functional to be real forces to use the natural involution \eqref{invol-4d} in the construction of $S_\kappa$\footnote{Notice that $S_\kappa$ describes {\it{a priori}} the dynamics of a complex-valued field (obvious from \eqref{invol-4d}) unless one imposes the additional constraint $\bar{\phi}=\phi$, which therefore would give rise to a NCFT for a real-valued field.}. Recall that this involution is rigidly linked to the construction of the C*-algebra modeling the $\kappa$-Minkowski space (see section \ref{section2-1}). Hence, according to the discussion given section \ref{section311} and eqn. \eqref{positivity}, admissible (positive) real quartic (star) polynomial interactions are of the form $\langle f,f\rangle$, with $f\in\mathcal{M}_\kappa$ a polynomial in the fields $\phi$ and $\phi^\dag$. They are given by
\begin{align}
\langle\phi^\dag\star\phi,\phi^\dag\star\phi\rangle,\ \langle\phi^\dag\star\phi^\dag,\phi^\dag\star\phi^\dag\rangle,\ \langle\phi\star\phi^\dag,\phi\star\phi^\dag\rangle,\ \langle\phi\star\phi,\phi\star\phi\rangle,
\end{align}
leading respectively to the following real interactions terms, $\lambda\in\mathbb{R}$,
\begin{align}
S_{1;\kappa}^\text{int}&=\lambda\int d^4x\left(\phi^\dag\star\phi\star\phi^\dag\star\phi\right)(x),\ S_{2;\kappa}^\text{int}=\lambda\int d^4x\left(\phi\star\phi\star\phi^\dag\star\phi^\dag\right)(x),\label{interac-type1}\\
S_{3;\kappa}^\text{int}&=\lambda\int d^4x\left(\phi\star\phi^\dag\star\phi\star\phi^\dag\right)(x),\ S_{4;\kappa}^\text{int}=\lambda\int d^4x\left(\phi^\dag\star\phi^\dag\star\phi\star\phi\right)(x).\label{interac-type2}
\end{align}
The existence of these four different families of interactions reflects the non-commutativity of the star product, as well as the non-cyclicity of the integral, involved in $S_\kappa$, although they all admit the same commutative limit $\lambda\vert\phi\vert^4$, eqn. \eqref{action-limit}. In fact, the second set of interactions \eqref{interac-type2} differs from the first one \eqref{interac-type1} by some power of the twist factor $\sigma$, eqns. \eqref{twistrace}, \eqref{twistoperator}, as it can be easily realised upon using \eqref{twistrace} in \eqref{interac-type2}\footnote{Straightforward application of the twisted trace property of the integral \eqref{twistrace} in \eqref{interac-type2} yields
\begin{equation}\label{interac-type2bis}
S_{3;\kappa}^\text{int}=\lambda\int d^4x\left((\sigma\triangleright\phi^\dag)\star\phi\star\phi^\dag\star\phi\right)(x),\ \ S_{4;\kappa}^\text{int}=\lambda\int d^4x\left((\sigma\triangleright(\phi\star\phi))\star\phi^\dag\star\phi^\dag\right)(x).
\end{equation}
\label{foot-twist}}. The actual non-equivalence of the four interactions becomes more apparent after using the integral expressions for the star product \eqref{starpro-4d} and involution \eqref{invol-4d} in \eqref{interac-type1} and \eqref{interac-type2}, leading to the expressions for the corresponding vertex-functions, eqns. \eqref{vertex-p1}-\eqref{vertex-p4}. Anticipating the results of the section \ref{subsection32}, it will be shown that each of these theories leads to different quantum (one-loop) corrections to the 2-point functions.\\
Notice that $S_{1;\kappa}^\text{int}$ and $S_{3;\kappa}^\text{int}$ (resp. $S_{2;\kappa}^\text{int}$ and $S_{4;\kappa}^\text{int}$) may be viewed as so-called orientable (resp. non-orientable) interaction, according to the standard liturgy of NCFT, each type leading to its own quantum behavior for the corresponding NCFT. For more technical details on the diagrammatic, see e.g \cite{vign-sym,thes-vt} for NCFT on Moyal space and \cite{poulwal-1} for the $\mathbb{R}^3_\theta$ case and references therein.\\
Notice also that these four interactions obviously reduce to a single one whenever $\phi$ satisfies $\phi^\dag=\phi$. The resulting interaction actually coincides with the quartic  interaction considered in \cite{GAC2002} only when the field $\phi$ satisfies the additional constraint $\bar{\phi}=\phi$, i.e $\phi$ is real-valued. This can be explicitly verified by standard computation from eqns. \eqref{vertex-p1}-\eqref{vertex-p4} given below. Recall that in \cite{GAC2002}, a nice use is made of path integral quantization methods to investigate some properties of real-valued scalar NCFT with quartic interaction on $\kappa$-Minkowski space, in particular the non-linear conservation law characterizing the interaction.\\

As we did for the kinetic term, it is convenient to express $S_{I;\kappa}^\text{int}$, $I=1,2,3,4$, as (commutative) non-local interaction terms involving $\phi$ and $\bar{\phi}$. This is achieved by successively using \eqref{algeb-1} and \eqref{starpro-4d}, \eqref{invol-4d} in \eqref{interac-type1} and \eqref{interac-type2}. Standard computations yield
\begin{equation}
S_{I;\kappa}^\text{int}=(2\pi)^4\lambda\int \left[\prod_{i=1}^4d^4x_i\right] \ \bar{\phi}(x_1) \phi(x_2) \bar{\phi}(x_3) \phi(x_4) \mathcal{V}_{I;\kappa}(x_1,x_2,x_3,x_4), \label{inter-real-gene}
\end{equation}
where the vertex function takes the form
\begin{equation}
\mathcal{V}_{I;\kappa}(x_1,x_2,x_3,x_4) = \int \left[\prod_{i=1}^4\frac{d^4p_i}{(2\pi)^4}\right] \ \widetilde{\mathcal{V}}_{I;\kappa}(p_1,p_2,p_3,p_4)e^{i(p_1\cdot x_1 - p_2\cdot x_2 + p_3\cdot x_3 - p_4\cdot x_4)}.\label{real-xvertex-gene}
\end{equation}
The explicit expressions for the vertex functions characterising the above interactions, \eqref{interac-type1} and \eqref{interac-type2}, are given by
\begin{align}
\widetilde{\mathcal{V}}_{1;\kappa}(\lbrace p_i\rbrace)& = \delta\left(p_2^0-p_1^0+p_4^0-p_3^0\right)\delta^{(3)}\left(\left(\vec{p}_2-\vec{p}_1\hspace{2pt}\right)e^{p_1^0/\kappa}+\left(\vec{p}_4-\vec{p}_3\hspace{2pt}\right)e^{p_4^0/\kappa}\right),\label{vertex-p1}\\
\widetilde{\mathcal{V}}_{2;\kappa}(\lbrace p_i\rbrace) &= \delta\left(p_2^0-p_1^0+p_4^0-p_3^0\right)\delta^{(3)}\left(\vec{p}_2-\vec{p}_1+\vec{p}_4e^{-p_2^0/\kappa}-\vec{p}_3e^{-p_1^0/\kappa}\right)\label{vertex-p2},\\
\widetilde{\mathcal{V}}_{3;\kappa}(\lbrace p_i\rbrace) &=\delta\left(p_2^0-p_1^0+p_4^0-p_3^0\right)\delta^{(3)}\left(\vec{p}_2-\vec{p}_3+\vec{p}_4e^{(p_4^0-p_3^0)/\kappa}-\vec{p}_1e^{(p_1^0-p_2^0)/\kappa}\right), \label{vertex-p3}\\
\widetilde{\mathcal{V}}_{4;\kappa}(\lbrace p_i\rbrace) &= \delta\left(p_2^0-p_1^0+p_4^0-p_3^0\right) \delta^{(3)}\left((\vec{p}_2+\vec{p}_4e^{-p_4^0/\kappa})e^{-p_2^0/\kappa}-(\vec{p}_1+\vec{p}_3e^{-p_3^0/\kappa})e^{-p_1^0/\kappa}\right)\label{vertex-p4}.
\end{align}

Equations \eqref{vertex-p1}-\eqref{vertex-p4} exhibit the energy-momentum conservation laws for each of those theories. As expected, the conservation law for the energy (time-like momenta) sector is the standard one while the 3-momentum conservation law becomes non linear. This stems from the semi-direct product structure underlying the non-commutative C*-algebra modelling $\kappa$-Minkowski and reflects the (Hopf algebraic) structure of the $\kappa$-Poincar\'e algebra underlying its (quantum) symmetries. Note this is sometimes geometrically interpreted (for instance in the context of relative locality, see e.g. \cite{reloc}) as reflecting the existence of a curvature of the energy-momentum space at very high (i.e. of order $\kappa$) energy.\bigskip

Finally, in view of the above discussions and the explicit expressions for the $\widetilde{\mathcal{V}}_{I;\kappa}$'s characterising the various models, one easily convinces ourself that it is not possible to reduce the four (tree level) vertex functions \eqref{vertex-p1}-\eqref{vertex-p4} into one unique vertex (involving a unique delta function). This is obvious when considering two theories of different nature (i.e. either orientable or non-orientable). On the other hand, $S_{1;\kappa}$ and $S_{3;\kappa}$ (resp. $S_{2;\kappa}$ and $S_{4;\kappa}$) differ from each other by some power of the twist factor (see e.g. footnote \ref{foot-twist}). Hence, there is a priori no reason for the different interactions to describe equivalent theories. Computations of first order corrections to the 2-point functions (reported section \ref{subsection32}) will show that the different models studied have indeed very different quantum behaviours, the twist playing an important role in their actual UV behaviour. In particular, UV/IR mixing shows up for non-orientable theories albeit absent in the orientable models.
\subsection{One-loop 2-point functions.}\label{subsection32}
In this section, we present the computation of the one-loop 2-point functions for each of the two field theories characterized respectively by the interaction terms $S_{1;\kappa}^\text{int}$ and $S_{2;\kappa}^\text{int}$, \eqref{interac-type1}, both with kinetic term corresponding to \eqref{kinetic}.  We have verified that the field theories corresponding to interaction terms $S_{3;\kappa}^\text{int}$ or $S_{4;\kappa}^\text{int}$, \eqref{interac-type2}, exhibit a similar behaviour regarding the structure of the contributions received by the 2-point functions and their respective UV and IR behaviours. Their analysis can be obtained from straightforward adaptations of the material presented below. Anticipating the results, we find that the 2-point function for each of the theories \eqref{interac-type1} receives 4 types of contributions, hereafter denoted by Type-I, Type-II, Type-III and Type-IV. Type-I contributions can be interpreted as standard planar contributions while Type-II and Type-III contributions can be viewed as planar contributions stemming from the fact that the Lebesgue integral involved in the action is a twisted trace. The Type-IV contributions can be viewed as non-planar contributions which exhibit UV/IR mixing. Changing the kinetic term \eqref{kinetic} to \eqref{Casimir} does not modify noticeably the conclusions on the UV and IR behaviour of the field theories. 
This will be discussed in the Section \ref{section4}.
\subsubsection{Preliminary considerations.}
To deal with the perturbative expansion, we follow the usual route used in (most of) the studies of NCFT, which we briefly recall now. Namely, first by making use of \eqref{starpro-4d} and \eqref{invol-4d}, the action functional $S_\kappa(\phi^\dag,\phi)$ involving star products is represented as an ordinary, albeit non local, action functional $S_\kappa(\bar{\phi},\phi)$ depending on $\phi$, $\bar{\phi}$ and the ordinary (commutative) product among functions, hence describing the dynamics of a complex scalar field. Accordingly, the perturbative expansion related to the NCFT is nothing but a usual perturbative expansion for an ordinary (complex scalar) field theory, stemming from the generating functional of the connected correlation functions 
\begin{equation}
W_{I}[\bar{J},J]:=\ln\left(\mathcal{Z}_{I}[\bar{J},J]\right)
\end{equation}
with
\begin{equation}
\mathcal{Z}_{I}[\bar{J},J]:=\int d\bar{\phi}d\phi \ e^{-S^\text{kin}_\kappa(\bar{\phi},\phi)-S^\text{int}_{I;\kappa}(\bar{\phi},\phi)+\int d^4x \ \bar{J}(x)\phi(x) + \int d^4x \ J(x)\bar{\phi}(x)}, \ I=1,2,
\end{equation}
in which the functional measure is merely the ordinary functional measure for a scalar field theory $S_\kappa(\bar{\phi},\phi)$ implementing formally the integration over the field configurations $\phi$ and $\bar{\phi}$. Accordingly, correlation functions built from $\phi$ and $\bar{\phi}$ are then generated by the repeated action of standard functional derivatives with respect to $J$ and $\bar{J}$ satisfying the usual functional rule
\begin{equation}
\frac{\delta J(p)}{\delta J(q)}=\delta^{(4)}(p-q).
\end{equation}
Note that there is no need to introduce a notion of non-commutative (star) functional derivative in the present approach. \bigskip

Let us recall, for the sake of completeness, the main steps of the derivation of the contributions to the one-loop 2-point functions. This can be achieved by first rewriting the interaction term $S^\text{int}_{I;\kappa}$ replacing the fields $\phi$ and $\bar{\phi}$ by the functional derivatives w.r.t. their corresponding sources $\bar{J}$ and $J$ respectively, then computing the Gaussian integral for the free field theory. This leads to
\begin{align}
&W_{0}[\bar{J},J]:=\int \frac{d^4p}{(2\pi)^4} \ \overline{\mathcal{F}J}(p)P_\kappa(p)\mathcal{F}J(p),\\
W_{I}[\bar{J},J]=\ln N&+W_{0}[\bar{J},J]+\ln\left(1+e^{-W_{0}}\left(e^{-S_{I;\kappa}^\text{int}[\frac{\delta}{\delta \mathcal{F}J},\frac{\delta}{\delta\overline{\mathcal{F}J}}]}-1\right)e^{W_{0}}\right),\label{expansion1}
\end{align}
with $N$ some normalisation constant, $P_\kappa(p)$ the Fourier transform of \eqref{propagator} and where we have switched from position to momentum representation for computational convenience.\\Now expanding the last logarithm in \eqref{expansion1} up to the first order in the coupling constant $\lambda$ and defining the effective action $\Gamma$ as the Legendre transform of $W_{I}$, 
\begin{equation}
\Gamma[\bar{\phi},\phi]:=\int \frac{d^4p}{(2\pi)^4} \left(\overline{\mathcal{F}J}(p)\mathcal{F}\phi(p)+\mathcal{F}J(p)\overline{\mathcal{F}\phi}(p)\right)-W_{I}[\bar{J},J],
\end{equation}
one finds, after standard computation, the following expression for the one-loop quadratic part of $\Gamma$
\begin{equation}
\Gamma^{(2)}_1[\bar{\phi},\phi]:=\int\frac{d^4p_3}{(2\pi)^4}\frac{d^4p_4}{(2\pi)^4}\  \overline{\mathcal{F}\phi}(p_3)\mathcal{F}\phi(p_4) \Gamma^{(2)}_1(p_3,p_4),
\end{equation}
with
\begin{align}
\Gamma^{(2)}_1(p_3,p_4):=\lambda\int &\frac{d^4p_1}{(2\pi)^4}\frac{d^4p_2}{(2\pi)^4}\ P_\kappa(p_1) \delta^{(4)}(p_2-p_1)\times\nonumber\\
&\times\Big[\widetilde{\mathcal{V}}_{I;\kappa}(p_1,p_2,p_3,p_4)+\widetilde{\mathcal{V}}_{I;\kappa}(p_3,p_4,p_1,p_2)+\nonumber\\
&\hspace{1cm}+\widetilde{\mathcal{V}}_{I;\kappa}(p_3,p_2,p_1,p_4)+\widetilde{\mathcal{V}}_{I;\kappa}(p_1,p_4,p_3,p_2)\Big],\label{gamma2}
\end{align}
The various contributions mentioned at the beginning of this section are then obtained by replacing $\widetilde{\mathcal{V}}_{I;\kappa}$ by the different expressions for the vertex function \eqref{vertex-p1}-\eqref{vertex-p4} in \eqref{gamma2}.
\subsubsection{Scalar theory with $\phi^\dag\star\phi\star\phi^\dag\star\phi$ interaction.}\label{321}
The relevant classical action functional is $S^\text{kin}_\kappa+S_{1;\kappa}^\text{int}$, see \eqref{pratik-kinet-act}, \eqref{pratik-kinet}, \eqref{interac-type1}. By a simple inspection of \eqref{gamma2}, one easily realizes that the one-loop 2-point function receives two types of contribution, hereafter called Type-I and Type-II contributions.\\
The contributions of Type-I are nothing but the usual planar contributions, according to the usual denomination prevailing in the non-commutative field theories. The Type-II contributions, while similar to the planar contributions in that they do not depend on the external momenta, are a new type of contributions generated by the twist which arises in the vertex functions, thus altering some diagrams with ``planar topology". No non-planar contributions (namely, depending on the external momenta) can be obtained within the present model so that no IR singularity related to the UV/IR mixing can occur in the 2-point function. Let us now study the UV behaviour of these contributions.\bigskip

Typical Type-I contribution to the one-loop effective action can be written as
\begin{equation}
\Gamma^{(2)}_{1;(I)}(p_3,p_4)= e^{-3p_3^0/\kappa}\delta^{(4)}(p_4-p_3)\Sigma_{(I)},\label{type1-struct}
\end{equation}
in which
\begin{equation} \label{Sigma1}
\Sigma_{(I)} := \lambda \int \frac{d^4p}{(2\pi)^4} \ \frac{e^{-2p^0/\kappa}}{1+e^{-3p^0/\kappa}} \  \frac{4\kappa^2}{\left(\vec{p}^{\hspace{2pt}2}+\kappa^2\mu^2_{+}\right)\left(\vec{p}^{\hspace{2pt}2}+\kappa^2\mu^2_{-}\right)}.
\end{equation}
Because of the strong decay of the propagator at large momentum $\vec{p}$ (, $\sim 1/\vec{p}^{\hspace{2pt}4}$), the spatial integral is finite and the integration over the 3-momentum $d^3\vec{p}$ can be performed by making use of the two following relations
\begin{align}
\frac{1}{A^aB^b}=\frac{\Gamma(a+b)}{\Gamma(a)\Gamma(b)} & \int_0^1 du \ \frac{u^{a-1}(1-u)^{b-1}}{\left(uA+(1-u)B\right)^{a+b}}, \ a,b>0, \\
\int \frac{d^np}{(2\pi)^n} \frac{1}{(p^2+M^2)^{m}}&=M^{n-2m}\frac{\Gamma(m-n/2)}{(4\pi)^{n/2} \Gamma(m)}, \ m>n/2> 0,
\end{align}
where $\Gamma(z)$ is the Euler gamma function. This leads to
\begin{equation}
\Sigma_{(I)} = \frac{2\kappa^2\lambda}{(2\pi)^2} \int_\mathbb{R} dp^0\ \frac{e^{-2p^0/\kappa}}{\left(1+e^{-3p^0/\kappa}\right)} \ \frac{\sqrt{\mu^2_{+}}-\sqrt{\mu^2_{-}}}{\mu^2_{+}-\mu^2_{-}},
\end{equation}
with $\mu^2_{+}-\mu^2_{-}=4\sqrt{1-\left(\frac{m}{\kappa}\right)^2}\ e^{-p^0/\kappa}$. By finally performing the change of variables, 
\begin{equation}
y=e^{-p^0/\kappa}, \label{change-var}
\end{equation}
$\Sigma_{(I)}$ reduces to
\begin{align}
&\Sigma_{(I)} = C \int_0^\infty dy \left[\frac{\sqrt{1+2\sqrt{1-\left(\frac{m}{\kappa}\right)^2}y+y^2}}{1+y^3}-\frac{\sqrt{1-2\sqrt{1-\left(\frac{m}{\kappa}\right)^2}y+y^2}}{1+y^3}\right], \label{typeI} \\
&\text{with}\ \ C:=\frac{\lambda}{(2\pi)^2}\frac{\kappa^3}{2\sqrt{\kappa^2-m^2}}, \label{constanteC}
\end{align}
whose UV behaviour can easily be inferred by use of the d'Alembert criterion as shown below. Before proceeding to that analysis, some comments are in order:
\begin{itemize}
\item First, notice that, due to the change of variables \eqref{change-var}, both the lower $(0)$ and upper $(\infty)$ bounds of integration  in \eqref{typeI} correspond to the UV (large $|p^0|$) regime. 
\item Next, some of the integrals w.r.t. the $y$ variable appearing in the computation of the one-loop order corrections to the 2-point function, have to be understood as regularized integrals. One way of regularizing them amounts to introduce a cut-off for $y$. Motivated by the Hopf algebraic structure of the $\kappa$-Poincar\'e algebra (in particular the deformed translation algebra), which is generated by $P_i$ and $\mathcal{E}=e^{-P_0/\kappa}$ (for more details see appendix \ref{apendixA}), it is natural to interpret $y=e^{-p^0/\kappa}$ as related to the ``physical" quantity replacing $p^0$ in the NCFT. More precisely, having in mind the expression for the 1st Casimir operator of the $\kappa$-Poincar\'e algebra, \eqref{Casimir2}, one can interpret the quantity
\begin{equation}
\mathcal{P}^0(\kappa):=\kappa(1-y)
\end{equation}
as the relevant quantity for the $\kappa$-Poincar\'e covariant quantum field theories, which reduces to $p^0$ when taking the formal commutative limit ($\kappa\to\infty$). Assuming $\vert\mathcal{P}^0\vert\leq\Lambda_0$,  it follows that one can derive an appropriate cut-off for $y$. This is achieved by noticing that the introduction of $\Lambda_0$ induces a cut-off for $p^0$, say $M_\kappa(\Lambda_0)$, which is easily shown to be related to $\Lambda_0$ by
\begin{equation}
M_\kappa(\Lambda_0) = \kappa \ln\left(1+\frac{\Lambda_0}{\kappa}\right),
\end{equation} 
with the limit $M_\kappa(\Lambda_0)\to\Lambda_0$ when $\kappa\to\infty$. Thus,
\begin{equation}
\frac{\kappa}{\kappa+\Lambda_0} \leq y \leq \frac{\kappa+\Lambda_0}{\kappa}.
\end{equation}
\end{itemize}
Having in mind these two comments, we can now study the UV behaviour of the scalar field theories under consideration.\\
When $y\to \infty$, one can check that 
\begin{equation}
\sqrt{\mu^2_{+}}-\sqrt{\mu^2_{-}}=2\sqrt{1-\left(\frac{m}{\kappa}\right)^2}+\mathcal{O}(\frac{1}{y^2}),\label{mumu}
\end{equation}
so that the integrand in eqn. \eqref{typeI} behaves like $\sim y^{-3}$. Meanwhile, when $y\to 0$, one verifies that the integrand behaves like $\sim y$. Hence, the integral is convergent, showing that typical Type-I contribution given by $\Sigma^{(I)}$ is (UV) finite. \bigskip

By performing a similar computation, one finds that typical contribution of Type-II have the same structure than those of Type-I \eqref{type1-struct}, still independent of external momenta, but receiving an extra contribution proportional to some power of $e^{-3p^0/\kappa}$ stemming from the twist $\sigma$, as indicated above. Indeed, the one-loop effective action can be cast into the form
\begin{equation}
\Gamma^{(2)}_{1;(I{\hspace{-2pt}I})}(p_3,p_4)=\delta^{(4)}(p_4-p_3)\Sigma_{(I{\hspace{-2pt}I})},
\end{equation}
where
\begin{equation}
\Sigma_{(I{\hspace{-2pt}I})} = \lambda \int \frac{d^4p}{(2\pi)^4} \ \frac{e^{-5p^0/\kappa}}{1+e^{-3p^0/\kappa}} \  \frac{4\kappa^2}{\left(\vec{p}^{\hspace{2pt}2}+\kappa^2\mu^2_{+}\right)\left(\vec{p}^{\hspace{2pt}2}+\kappa^2\mu^2_{-}\right)}. \label{Sigma2}
\end{equation}
Observe from \eqref{Sigma2} and \eqref{Sigma1} that one has formally 
\begin{equation}
\Sigma_{(I)}=\int \frac{d^4p}{(2\pi)^4}\ \mathcal{I}(p),\label{type1-sig}
\end{equation}
where the integrand $\mathcal{I}$ can be read off from \eqref{Sigma1}, while 
\begin{equation}
\Sigma_{(I{\hspace{-2pt}I})}=\int \frac{d^4p}{(2\pi)^4}\ e^{-3p^0/\kappa}\mathcal{I}(p),\label{type2-sig}
\end{equation}
in which the extra factor $e^{-3p^0/\kappa}$ is generated by a twist factor.\bigskip

Now, performing the change of variable \eqref{change-var} in \eqref{Sigma2} and characterizing the UV (large $|p^0|$) regime as done above for the Type-I contributions, one easily finds that the integral in \eqref{Sigma2} reduces to
\begin{equation}
\Sigma_{(I{\hspace{-2pt}I})} = C \int_0^\infty dy \ y^3 \left[\frac{\sqrt{1+2\sqrt{1-\left(\frac{m}{\kappa}\right)^2}y+y^2}}{1+y^3}-\frac{\sqrt{1-2\sqrt{1-\left(\frac{m}{\kappa}\right)^2}y+y^2}}{1+y^3}\right],
\end{equation}
where the constant $C$ is given by \eqref{constanteC}. The integral is still convergent for $y\to 0$ since the twist contributes by a factor $y^3$ at the numerator while the integrand behaves now like \eqref{mumu} when $y\to\infty$, instead of the convergent behaviour of the Type-I contribution. Hence, 
\begin{equation}
\Sigma_{(I{\hspace{-2pt}I})} \sim \frac{\lambda\kappa}{(2\pi)^2} \ \Lambda_0 + \lbrace \text{finite terms}\rbrace,
\end{equation}
which exhibits a linear UV divergence essentially produced by the twist in view of \eqref{type1-sig} and \eqref{type2-sig}.\bigskip

To summarize the results, we have found that within the field theory described by the action functional $S^\text{kin}_\kappa+S_{1;\kappa}^\text{int}$, the twist splits the planar contributions to the 2-point function into two different planar-like contributions which are IR finite and whose UV behaviour is affected by the twist. Note that all the contributions to the 2-point functions are independent of the external momenta so that no IR singularities at exceptional (zero) momenta, related to UV/IR mixing, can occur (there is no non-planar contributions).

\subsubsection{Scalar theory with $\phi\star\phi\star\phi^\dag\star\phi^\dag$ interaction.}\label{322}

The relevant classical action functional is now $S^\text{kin}_\kappa+S_{2;\kappa}^\text{int}$, see \eqref{pratik-kinet-act}, \eqref{pratik-kinet}, \eqref{interac-type1}. From the perturbative expansion of the corresponding partition function, one finds that the one-loop 2-point function receives three types of contribution, hereafter called Type-I, Type-III and Type-IV contributions.\\
The Type-I and Type-III are planar type, i.e. independent of the external momenta but differing from each other by its own contribution coming from the twist $\sigma$. This results in different powers of the factor $e^{-3p^0/\kappa}$ in the integrands of the various contributions, hence the denomination ``Type-III" since this factor is different from the one for Type-II contributions exhibited in the subsection \ref{321}. As for the field theory examined in subsection \ref{321}, Type-I contributions are found to be UV finite. The Type-IV contributions can be actually interpreted as non-planar contributions. This signals that the corresponding field theory has UV/IR mixing since Type-IV contributions evaluated at exceptional zero external momentum are divergent.\bigskip

Let us start by considering planar contributions. Typical Type-III contribution to the one-loop effective action can be written as
\begin{equation}
\Gamma^{(2)}_{1;(I{\hspace{-2pt}I}{\hspace{-2pt}I})}(p_3,p_4)= \delta^{(4)}(p_4-p_3)\Sigma_{(I{\hspace{-2pt}I}{\hspace{-2pt}I})},
\end{equation}
in which $\Sigma_{(I{\hspace{-2pt}I}{\hspace{-2pt}I})}=\int d^4p\ e^{3p^0/\kappa}\mathcal{I}(p)$, with $\mathcal{I}(p)$ defined in \eqref{type1-sig}. After performing the integration over $d^3\vec{p}$ and the change of variable \eqref{change-var}, one obtains 
\begin{equation}
\Sigma_{(I{\hspace{-2pt}I}{\hspace{-2pt}I})} =C \int_0^\infty \frac{dy}{y^3} \left[\frac{\sqrt{1+2\sqrt{1-\left(\frac{m}{\kappa}\right)^2}y+y^2}}{1+y^3}-\frac{\sqrt{1-2\sqrt{1-\left(\frac{m}{\kappa}\right)^2}y+y^2}}{1+y^3}\right].\label{type3-div}
\end{equation}
Using \eqref{mumu}, one easily finds that the integrand in \eqref{type3-div} behaves like $\sim y^{-6}$ when $y\to\infty$ while it behaves like $\sim y^{-2}$ for $y\to0$, such that
\begin{equation}
\Sigma_{(I{\hspace{-2pt}I}{\hspace{-2pt}I})}\sim \frac{\lambda\kappa}{(2\pi)^2} \ \Lambda_0 + \lbrace \text{finite terms}\rbrace ,
\end{equation}
indicating that \eqref{type3-div} has a UV linear divergence (as for Type-II contribution of the field theory considered in the previous subsection).\bigskip

Finally, let us consider the non-planar Type-IV contributions. That latter can be written as
\begin{equation}
\Gamma^{(2)}_{1;(I{\hspace{-2pt}V})}(p_3,p_4)=\delta(p_4^0-p_3^0)\Sigma_{(I{\hspace{-2pt}V})}(p_3,p_4),\label{decadix1}
\end{equation}
where
\begin{equation}
\Sigma_{(I{\hspace{-2pt}V})}(p_3,p_4) =  (2\kappa)^2\lambda \int \frac{d^4p}{(2\pi)^4} \ \frac{e^{-2p^0/\kappa}}{1+e^{-3p^0/\kappa}} \frac{\delta^{(3)}\left(\left(1-e^{-p_3^0/\kappa}\right)\vec{p}+\vec{p}_4e^{-p^0/\kappa}-\vec{p}_3\right)}{\left(\vec{p}^{\hspace{2pt}2}+\kappa^2\mu^2_{+}\right)\left(\vec{p}^{\hspace{2pt}2}+\kappa^2\mu^2_{-}\right)}. \label{nonplanar}
\end{equation}
Note that $\Sigma_{(I{\hspace{-2pt}V})}(p_3,p_4)$ depends on two (external) momenta which however are not independent, due to the (non-linear) momentum conservation ensured by the delta functions. This dependence by the way signals that the effective action functional \eqref{decadix1} is non-local.

Let's first examine the infrared sector. Setting $(p^0_3,\vec{p}_3)\to(0,\vec{0})$ in \eqref{nonplanar} leads to
\begin{equation}
\Sigma_{(I{\hspace{-2pt}V})}(0,p_4) = (2\kappa)^2\lambda \int \frac{d^4p}{(2\pi)^4} \ \frac{e^{-2p^0/\kappa}}{1+e^{-3p^0/\kappa}}\frac{e^{3p^0/\kappa}}{\left(\vec{p}^{\hspace{2pt}2}+\kappa^2\mu^2_{+}\right)\left(\vec{p}^{\hspace{2pt}2}+\kappa^2\mu^2_{-}\right)} \  \delta^{(3)}(\vec{p}_4),
\end{equation} 
such that
\begin{equation} \label{mixingIII}
\Sigma_{(I{\hspace{-2pt}V})}(0,p_4)=\delta^{(3)}(\vec{p}_4)\Sigma_{(I{\hspace{-2pt}I}{\hspace{-2pt}I})},
\end{equation}
indicating that the conservation law is preserved, namely $p_4\to 0$ when $p_3\to 0$, and that the non-planar contributions tends toward (Type-III) planar contributions in the limit of vanishing external momenta.

To study the UV behaviour of \eqref{nonplanar}, we perform the integration over $d^3\vec{p}$ together with the change of variables \eqref{change-var}. Standard computation yield
\begin{equation}\label{typeIV}
\Sigma_{(I{\hspace{-2pt}V})}(p_3,p_4) =  \frac{\kappa^2\lambda}{4\pi^4} \left|1-e^{-p_3^0/\kappa}\right| \int_0^\infty dy \ \frac{y}{\left(1+y^3\right)\Omega_{+}(y)\Omega_{-}(y)},
\end{equation}
with
\begin{equation}
\Omega_{\pm}(y) = \left(y\vec{p}_4-\vec{p}_3\right)^2+\kappa^2\left(1-e^{-p_3^0/\kappa}\right)^2\mu_{\pm}^2(y).
\end{equation}
Now, one can easily check that the integrand in \eqref{typeIV} behaves like $\sim y$ when $y\to0$, while it behaves like $\sim y^{-6}$ when $y\to\infty$.

Therefore, one concludes that Type-IV contributions are finite for any (non zero) external 4-momenta while $\lim_{p_3\to 0}\Sigma_{(I{\hspace{-2pt}V})}(p_3,p_4)\sim-\lambda\kappa\Lambda_0$, namely diverges (UV) linearly. This last phenomenon reflects the existence of perturbative UV/IR mixing when considering interactions of the form of $S^\text{int}_{2;\kappa}$. The same result occurs for interactions given by $S^\text{int}_{4;\kappa}$.

\section{Discussion and conclusion.}\label{section4}

The Weyl quantization scheme provides a natural framework to describe $\kappa$-deformations of the Minkowski space-time. A well controlled star product for $\kappa$-Minkowski space is easily obtained from the representations of the convolution algebra of the affine group which here replaces the Heisenberg group underlying the popular quantization of a phase space. Owing to the fact that the $\kappa$-Minkowski space supports a natural action of a deformation of the Poincar\'e Lie algebra, the $\kappa$-Poincar\'e algebra playing the role of the algebra of symmetry of the quantum space, it is physically relevant to require $\kappa$-Poincar\'e invariance of any physically reasonable action functional. Doing this necessarily implies that the trace building the action functional is twisted, stemming simply from the peculiar behavior of the star product w.r.t. the Lebesgue integral involved in the action.\bigskip

We have examined various classes of (complex) scalar field theories on 4-d $\kappa$-Minkowski space, considering all possible types of quartic interaction allowed by reality condition of the action functional, and whose commutative limit coincides with the standard (commutative) complex $\phi^4$ theory. The kinetic operators were chosen to be square of different Dirac operators. The use of algebraic properties of the twisted trace leads to an easy computation of the corresponding propagators, despite the fact that the star product is not closed w.r.t. the integral.\bigskip

Focusing first on a kinetic operator \eqref{kinetic} related to the Dirac operator of an equivariant spectral triple considered in \cite{frans-2}, we have analyzed the one-loop UV and IR behavior of the 2-point functions for each of these theories, presenting in details the technical analysis for representative classes of theories \eqref{interac-type1} in the subsection \ref{subsection32}. We find that the twist splits the planar contributions to the 2-point function into different {\it{IR finite}} contributions whose UV behavior depends on the power of the twist factor arising, technically speaking, from the respective positions of the contracted fields in the interaction combined with the non-cyclicity of the trace. The UV behavior of these contributions ranges from UV finitude to at most UV linear divergence, which is slightly milder than in the commutative scalar theory. The interaction term of the scalar theory considered in the subsection \ref{321} cannot produce non-planar contributions, since the interaction is orientable (in the terminology of non-commutative field theories). Hence, no UV/IR mixing is expected to occur in this field theory which therefore should be perturbatively renormalizable to all orders.\\
It turns out that the computation of the 1-loop contributions to the 4-point function for this NCFT shows that this latter is UV finite. The full derivation is cumbersome and will be reported elsewhere \cite{PW-1} together with the analysis of 2- and 4-point functions for the other NCFT considered in this paper. The UV finiteness is partly due to the large spatial momentum behavior of the propagator which decays as $1/\vec{p}^{\ 4}$. This yields finite spatial integrals for all the contributions while each of the remaining integrals over $y$ is found to be finite by a mere use of d'Alembert criterion. This additional observation together with the strong decay of the propagator at large (spatial) momenta makes very likely the perturbative renormalisability of this NCFT to all orders.\bigskip

UV/IR mixing is expected to occur in the scalar theory of subsection \ref{322} (the interaction is no longer orientable). Indeed, we find that the so-called Type-IV contribution, which depends on the external momenta, is finite for non zero external moment while it becomes singular at exceptional zero external momenta, see for instance \eqref{mixingIII}. It would be interesting to examine if this UV/IR mixing could be removed by using procedures similar to the one used to deal with the mixing within non-commutative field theories 
on Moyal spaces \cite{Grosse:2003aj-pc}. The above conclusions apply to the 2-point functions of the field theories \eqref{interac-type2}, whose analysis can be obtained from straightforward adaptations of subsection \ref{subsection32}. In the same way, changing the kinetic term \eqref{kinetic} to \eqref{Casimir} does not modify significantly the conclusions on the UV and IR behavior of these field theories. For instance, for the theory considered section \ref{321}, the Type-I contribution remains finite whereas the Type-II contribution diverges quadratically. Note that our conclusions qualitatively agree with those obtained a long time ago in \cite{gross-whl} where a scalar field theory built from another (albeit presumably equivalent) star product and a different kinetic operator has been considered. Again, linear UV divergences for planar-type contributions together with UV/IR mixing in non-planar contributions was shown to occur in that model. The precise comparison between both work is however drastically complicated by the technical approach used in \cite{gross-whl} leading to very involved formulas.\bigskip

An immediate natural extension of this analysis is the computation of the one-loop corrections to the vertex functions and beta functions in the above field theories. The corresponding work will be reported elsewhere \cite{PW-1}. The extension of the present work to the case of gauge theories defined on $\kappa$-Minkowski spaces is an interesting issue \cite{PSW-1}. In view of the natural action of the $\kappa$-Poincar\'e algebra, the framework of bicovariant differential calculus \cite{Maj-sit} seems to be better suited here than the standard derivation-based differential calculus  with which most of the non-commutative gauge models on $\mathbb{R}^4_\theta$ or $\mathbb{R}^3_\lambda$ have been built \cite{mdv-jcw}. A suitable framework should presumably take into account algebras of twisted derivations as well as twisted gauge transformations.\bigskip 

To conclude, we mention that the star product considered in this paper could be used in the construction of other (even non $\kappa$-Poincar\'e invariant) NCFT or gauge versions of them and should prove convenient to compute related quantum corrections. We note that the NCFT with orientable interaction \eqref{vertex-p1} provides an explicit example (as far as we know the first one) of a UV/IR mixing free NCFT on the 4-d $\kappa$-Minkowski space which is very likely renormalisable to all orders. It would be very interesting to show if some KMS condition stemming from the twisted trace rules the correlation functions of this NCFT which would signal the appearance of an observer-independent time within this theory and would then give to the NCFT on $\kappa$-Minkowski space a new impulse toward potential applications to fundamental physics.

\vskip 0,5 true cm
{\bf{Acknowledgments:}} J.-C. Wallet thanks N. Franco for discussions related to the present work. T. Poulain is grateful to A. Sitarz for discussions on material of ref. \cite{DS}.

\appendix
\section{Basics on $\kappa$-Poincar\'e algebra and deformed translations.}\label{apendixA}

Let $\mathcal{P}_\kappa$ denote the $\kappa$-Poincar\'e algebra. Let $\Delta:\mathcal{P}_\kappa\otimes\mathcal{P}_\kappa\to\mathcal{P}_\kappa$, $\epsilon:\mathcal{P}_\kappa\to\mathbb{C}$ and $S:\mathcal{P}_\kappa\to\mathcal{P}_\kappa$ be respectively the coproduct, counit and antipode, thus endowing $\mathcal{P}_\kappa$ with a Hopf algebra structure. A convenient presentation of $\mathcal{P}_\kappa$ is obtained from the 11 elements $(P_i, N_i,M_i, \mathcal{E},\mathcal{E}^{-1})$, $i=1,2,3$, respectively the momenta, the boost, the rotations and $\mathcal{E}:=e^{-P_0/\kappa}$ satisfying the Lie algebra relations{\footnote{In the following, Greek (resp. Latin) indices label as usual space-time (resp. purely spatial) coordinates.}}
\begin{equation}
[M_i,M_j]= i\epsilon_{ij}^{\hspace{5pt}k}M_k,\ [M_i,N_j]=i\epsilon_{ij}^{\hspace{5pt}k}N_k,\ [N_i,N_j]=-i\epsilon_{ij}^{\hspace{5pt}k}M_k\label{poinc1}, 
\end{equation}
\begin{equation}
[M_i,P_j]= i\epsilon_{ij}^{\hspace{5pt}k}P_k,\ [P_i,\mathcal{E}]=[M_i,\mathcal{E}]=0,\ [N_i,\mathcal{E}]=-\frac{i}{\kappa}P_i\mathcal{E}\label{poinc2},
\end{equation}
\begin{equation}
[N_i,P_j]=-\frac{i}{2}\delta_{ij}\left(\kappa(1-\mathcal{E}^{2})+\frac{1}{\kappa}\vec{P}^2\right)+\frac{i}{\kappa}P_iP_j\label{poinc3}, 
\end{equation}
with the Hopf algebra structure defined by
\begin{align}
\Delta P_0&=P_0\otimes\bbone+\bbone\otimes P_0,\ \Delta P_i=P_i\otimes\bbone+\mathcal{E}\otimes P_i,\label{hopf1}\\
\Delta \mathcal{E}&=\mathcal{E}\otimes\mathcal{E},\ \Delta M_i=M_i\otimes\bbone+\bbone\otimes M_i,\label{hopf1bis}\\
\Delta N_i&=N_i\otimes \bbone+\mathcal{E}\otimes N_i-\frac{1}{\kappa}\epsilon_{i}^{\hspace{2pt}jk}P_j\otimes M_k,\label{hopf2}
\end{align}
and 
\begin{align}
\epsilon(P_0)&=\epsilon(P_i)=\epsilon(M_i)=\epsilon(N_i)=0,\  \epsilon(\mathcal{E})=1\label{hopf3},\\
S(P_0)&=-P_0,\ S(\mathcal{E})=\mathcal{E}^{-1},\  S(P_i)=-\mathcal{E}^{-1}P_i,\label{hopf4}\\
S(M_i)&=-M_i,\ S(N_i)=-\mathcal{E}^{-1}(N_i-\frac{1}{\kappa}\epsilon_{i}^{\hspace{2pt}jk}P_jM_k)\label{hopf4bis}.
\end{align}
Recall that the $\kappa$-Minkowski space can be viewed as the dual of the Hopf subalgebra generated by $P_\mu$, $\mathcal{E}$, sometimes called the ``deformed translation algebra". This latter becomes a $^*$-Hopf algebra through: $P_\mu^\dag=P_\mu$, $\mathcal{E}^\dag=\mathcal{E}$. Then, by promoting the above duality to a duality between $^*$-algebras insuring compatibility among the involutions, one obtains
\begin{equation}
(t\triangleright f)^\dag=S(t)^\dag\triangleright f,\label{pairing-involution}
\end{equation}
which holds true for any $t$ in the deformed translation algebra and for any $f\in\mathcal{M}_\kappa$. This, combined with \eqref{hopf4} implies
\begin{equation}
(P_0\triangleright f)^\dag=-P_0\triangleright(f^\dag),\ (P_i\triangleright f)^\dag=-\mathcal{E}^{-1}P_i\triangleright(f^\dag),\ (\mathcal{E}\triangleright f)^\dag=\mathcal{E}^{-1}\triangleright(f^\dag)\label{dag-hopfoperat}.
\end{equation}
It must be stressed that the $P_i$'s act as twisted derivations on $\mathcal{M}_\kappa$ while $P_0$ remains untwisted as it can be readily seen from \eqref{hopf1}. One has for any $f,g\in\mathcal{M}_\kappa$
\begin{align}
P_i\triangleright(f\star g)&=(P_i\triangleright f)\star g+(\mathcal{E}\triangleright f)\star (P_i\triangleright g)\label{deriv-twist1},\\
P_0\triangleright(f\star g)&=(P_0\triangleright f)\star g+f\star(P_0\triangleright  g )\label{deriv-twist2}.
\end{align}
Note that $\mathcal{E}$ is not a derivation of $\mathcal{M}_\kappa$ since one has
\begin{equation}
\mathcal{E}\triangleright(f\star g)=(\mathcal{E}\triangleright f)\star(\mathcal{E}\triangleright g).\label{relation-calE}
\end{equation}

The structure of $\mathcal{M}_\kappa$ as left-module over the Hopf algebra $\mathcal{P}_\kappa$ can be expressed, for any $f\in\mathcal{F}(\mathcal{S}_c)$, in terms of the bicrossproduct basis $(M_i,N_i,P_\mu)$, \cite{majid-ruegg}, by
\begin{align}
(\mathcal{E}\triangleright f)(x)&=f(x_0+\frac{i}{\kappa},\vec{x})\label{left-module0},\\
(P_\mu\triangleright f)(x)&=-i(\partial_\mu f)(x),\label{left-module1}\\
(M_i\triangleright f)(x)&=\left(\epsilon_{ijk}L_{x_j}P_k\triangleright f\right)(x),\label{left-modules1bis}\\
(N_i\triangleright f)(x)&=\bigg(\big(\frac{1}{2}L_{x_i}(\kappa(1-\mathcal{E}^2)+\frac{1}{\kappa}\vec{P}^{2})+L_{x_0}P_i-\frac{i}{\kappa}L_{x_k}P_kP_i   \big)\triangleright f\bigg)(x),\label{left-module2}
\end{align}
where $L_a$ denotes the left (standard) multiplication operator, i.e. $L_af:=af$.
\section{KMS weight and twisted trace.}\label{apendixB}
A KMS weight on a (C*-)algebra $\mathbb{A}$ for a modular group of $^*$-automorphisms $\{\sigma_t\}_{t\in\mathbb{R}}$ is defined \cite{kuster} as a (densely defined) linear 
map $\varphi:\mathbb{A}_+\to\mathbb{R}^+$ ($\mathbb{A}_+$ is the set of positive elements of $\mathbb{A}$) such that $\{\sigma_t\}_{t\in\mathbb{R}}$ admits an analytic extension, still a one-parameter group, $\{\sigma_z\}_{z\in\mathbb{C}}$ acting on $\mathbb{A}$ satisfying the following two conditions{\footnote{Some alternative equivalent definitions exist, which however are less convenient for the present discussion. The above definition \cite{kuster} also require that $\varphi$ is lower semi-continuous and that $\{\sigma_z\}$ is norm-continuous, two conditions which are fortunately fulfilled in this paper.}}: 
\begin{equation}
{\textrm{i)}}\ \ \varphi\circ\sigma_z=\varphi,\ \ {\textrm{ii)}}\ \ \varphi(a^\dag \star a)=\varphi(\sigma_{\frac{i}{2}}(a)\star(\sigma_{\frac{i}{2}}(a))^\dag),\label{prop-kmsweight}
\end{equation}
for any $a$ in the domain of $\sigma_{\frac{i}{2}}$. The notion of weight on a C*-algebra extends the usual notion of state, since a state can be viewed (up to technical subtleties) as a weight with unit norm. In the present situation, the characterization of the relevant C*-algebra has been discussed in the section \ref{section2}. For our purpose, it will be sufficient to keep in mind that it involves $\mathcal{M}_\kappa$ as a dense $^*$-subalgebra. For more mathematical details on KMS weights, see e.g \cite{kuster}. Note that the notion of KMS weight related to the present twisted trace has been already used in \cite{matas} to construct a modular spectral triple for $\kappa$-Minkowski space.\bigskip

To verify that the twisted trace \eqref{twistrace}, \eqref{twistoperator} is actually a KMS weight, we first characterize the properties of $\sigma_t$ \eqref{sigmat-modul}. From \eqref{sigmat-modul}, \eqref{starpro-4d} and \eqref{invol-4d}, one obtains
\begin{equation}
\sigma_{t_1}\sigma_{t_2}=\sigma_{t_1+t_2},\ \sigma^{-1}_t=\sigma_{-t},\ \ \forall t,t_1,t_2\in\mathbb{R}, \label{modulargroup}
\end{equation}
and 
\begin{equation}
\sigma_t(f\star g)=\sigma_t(f)\star\sigma_t(g),\ \ \sigma_t(f^\dag)=(\sigma_t(f))^\dag,\ \forall t\in\mathbb{R}\label{modular-sigma},
\end{equation}
for any $f,g\in\mathcal{M}_\kappa$. Hence $\sigma_t$ \eqref{sigmat-modul} defines a group of $^*$-automorphisms of $\mathcal{M}_\kappa$. 
Next, set $\varphi(f):=\int d^4x\ f(x)$. Then, $\varphi$ verifies the property i) of \eqref{prop-kmsweight} as a mere consequence of \eqref{invarquant}, i.e. the $\kappa$-Poincar\'e invariance of the action functional. Namely
\begin{equation}
\varphi(\sigma_t f)=\sigma_t\triangleright\int d^4x\ f(x)=(\mathcal{E})^{-i3t}\triangleright\int d^4x\ f(x)=\epsilon(\mathcal{E})^{-i3t}\int d^4x\ f(x)=\varphi(f),
\end{equation}
for any $f\in\mathcal{M}_\kappa$ where the action of $\mathcal{E}$ has been extended to the one of $\sigma_t$ by using the functional calculus. \\
Before we verify the property ii) of \eqref{prop-kmsweight}, one remark is in order. Extend $\sigma_t$ \eqref{modular-sigma} to
\begin{equation}
\sigma_z(f):=e^{iz\frac{3P_0}{\kappa}}\triangleright f=e^{\frac{3z}{\kappa}\partial_0}\triangleright f,\ \forall z\in\mathbb{C}\label{sigmat-modul-z},
\end{equation}
for any $f\in\mathcal{M}_\kappa$. Then, one can easily verify that \eqref{modulargroup} and \eqref{modular-sigma} extend respectively to
\begin{equation}
\sigma_{z_1}\sigma_{z_2}=\sigma_{z_1+z_2},\ \sigma^{-1}_z=\sigma_{-z},\ \ \forall z,z_1,z_2\in\mathbb{C}, \label{prop-modulargroup-z}
\end{equation}
and 
\begin{equation}
\sigma_z(f\star g)=\sigma_z(f)\star\sigma_z(g)\label{morphalg-modul-z},
\end{equation}
while $\sigma_z$ is no longer an automorphism of $^*$-algebra. Namely, one has
\begin{equation}
\sigma_z(f^\dag)=(\sigma_{\bar{z}}(f))^\dag,\ \forall z\in\mathbb{C}\label{modulargroup-z}.
\end{equation}
In particular, the twist $\sigma$ \eqref{twistoperator} is recovered for $z=i$, i.e.
\begin{equation}
\sigma=\sigma_{z=i}\label{b10}
\end{equation}
and one has $\sigma(f^\dag)=(\sigma^{-1}(f))^\dag$. This type of automorphim is known as a regular automorphim in the mathematical literature and occurs in the framework of twisted spectral triples. It has been introduced in \cite{como-1} in conjunction with the assumption of the existence of a distinguished group of ($^*$-)automorphisms of the algebra indexed by one real parameter, says $t$, i.e. the modular group, such that the analytic extension $\sigma_{t=i}$ coincides precisely with the regular automorphism. Here, the modular group linked with the twisted trace is defined by $(\sigma_t)_{t\in\mathbb{R}}$ while the twist $\sigma=\sigma_{t=i}$ defines the related regular automorphism.\bigskip

To verify the 2nd property of \eqref{prop-kmsweight}, we use \eqref{morphalg-modul-z}, \eqref{modulargroup-z} and \eqref{twistrace} to compute the RHS of ii) \eqref{prop-kmsweight}. One has
\begin{align}
\varphi(\sigma_{\frac{i}{2}}(f)\star(\sigma_{\frac{i}{2}}(f))^\dag)&=\int d^4x\ \sigma_{\frac{i}{2}}(f)\star\sigma_{-\frac{i}{2}}(f^\dag)=\int d^4x\ \sigma_{\frac{i}{2}}(f\star\sigma_{-i}(f^\dag))\nonumber\\
&=\int d^4x\ f\star\sigma_{-i}(f^\dag)=\int d^4x\ \sigma(\sigma_{-i}(f^\dag))\star f
=\varphi(f^\dag\star f)\label{proof-kmsweight}.
\end{align} 
Hence $\varphi(f)=\int d^4x\ f(x)$ for any $f\in\mathcal{M}_\kappa$ defines a KMS weight. \bigskip

Now, the Theorem [6.36] of the 1st of ref. \cite{kuster} guarantees, for each pair $(a,b)\in\mathbb{A}$, the existence of a bounded continuous function $f:\Sigma\to\mathbb{C}$, where $\Sigma$ is the strip defined by $\{z\in\mathbb{C},\ 0\le\textrm{Im}(z)\le 1\}$, such that one has
\begin{equation}
f(t)=\varphi(\sigma_t(a)\star b),\ \ f(t+i)=\varphi(b\star\sigma_t(a)),\label{KMS-abst}
\end{equation}
where it is easy to realize that eqn. \eqref{KMS-abst} is an abstract version of the KMS condition.\bigskip

Note that $\sigma_t$ \eqref{sigmat-modul} defines ``time translations" since one has $\sigma_t(\phi)(x_0,\vec{x})=\phi(x_0+\frac{3t}{\kappa},\vec{x})$. Now we introduce the GNS representation of $\mathcal{M}_\kappa$, $\pi_{GNS}:\mathcal{F}(\mathcal{S}_c)\to\mathcal{B}(\mathcal{H})$, defined as usual by $\pi_{GNS}(\phi)\cdot v=\phi\star v$ for any $v\in\mathcal{H}$ where $\mathcal{H}$ is the Hilbert space unitary equivalent to $L^2(\mathbb{R}^4)$ discussed in the subsection \ref{section31}. Then, we compute 
\begin{eqnarray}
\pi_{GNS}(\sigma_t\phi)\cdot\omega&=&(\sigma_t\phi)\star\omega=\sigma_t(\phi\star(\sigma_t^{-1}\omega))
=(\sigma_t\odot\pi_{GNS}(\phi)\odot\sigma_t^{-1})\cdot(\omega)\nonumber\\
&=&((\Delta_T)^{it}\odot\pi_{GNS}(\phi)\odot(\Delta_T)^{-it})\cdot\omega\label{compo2},
\end{eqnarray}
for any $\omega\in\mathcal{H}$ and any $\phi\in\mathcal{M}_\kappa$, where $\odot$ stands for the composition law of maps (not to be confused with the convolution product) and $\Delta_T$ is the Tomita operator given by
\begin{equation}
\Delta_T=e^{\frac{3P_0}{\kappa}},
\end{equation}
which coincides with \eqref{modul-4d} and such that $\sigma_t=(\Delta_T)^{it}$. Eqn. \eqref{compo2} indicates that the modular group defined by $\{\sigma_t\}_{t\in\mathbb{R}}$ generates a ``temporal" evolution for the operators stemming from the Weyl quantization map $Q$. \bigskip

\section{Characterization of the Hilbert space.}\label{apendixC}

The Hilbert space $\mathcal{H}$ related to the Hilbert product \eqref{hilbert-product} can be obtained canonically from the GNS construction by completing the linear space $\mathcal{F}(\mathcal{S}_c)$ with respect to the natural norm
\begin{equation}
||f||^2=\langle f,f\rangle=\int d^4x\ \left(f^\dag\star f\right)(x)=\int d^4x\ |f^\dag(x)|^2.
\end{equation}
Unitary equivalence between $\mathcal{H}$ and $L^2(\mathbb{R}^4)$ can be easily shown by considering the (invertible) intertwiner map $A_\kappa:\mathcal{F}(\mathcal{S}_c)\to L^2(\mathbb{R}^4)$ which is defined for any $f\in\mathcal{F}(\mathcal{S}_c)$ by 
\begin{equation}
(A_\kappa f)(x)=\int \frac{dp^0}{2\pi} dy_0\ e^{iy_0p^0}{f}(x_0+y_0,e^{-p^0/\kappa}\vec{x}),\label{intertwin}
\end{equation}
with 
\begin{equation}
{\overline{(A_\kappa f)}}(x)=f^\dag(x)\label{conjug-unitar}.
\end{equation}
It follows immediately that $||A_\kappa f||^2_2=\int d^4x\ {\overline{(A_\kappa f)}}(x)(A_\kappa f)(x)=\int d^4x\ |f^\dag(x)|^2=||f||^2$. Therefore $A_\kappa $ defines an isometry which, owing to the density of $\mathcal{F}(\mathcal{S}_c)$ in $\mathcal{H}$, extends to $\mathcal{H}\to L^2(\mathbb{R}^4)$ while surjectivity of $A_\kappa$ stems directly from the existence of $A_\kappa^{-1}$ together with density of $\mathcal{F}(\mathcal{S}_c)$ in $L^2(\mathbb{R}^4)$. This proves that $A_\kappa$ is unitary together with the unitary equivalence mentioned above. Note that one verifies that $A_\kappa^{-1}$ is simply given by
\begin{equation}
(A_\kappa^{-1}f)(x)=\int \frac{dp^0}{2\pi} dy_0e^{-ip^0y_0}f(x_0+y_0,e^{-p^0/\kappa}\vec{x}),\label{intertwin-invers}
\end{equation}
for any $f\in\mathcal{F}(\mathcal{S}_c)$ so that \eqref{conjug-unitar} takes the convenient form $f^\dag(x)=(A_\kappa^{-1}\bar{f})(x)$.


\begin{thebibliography}{10}
\bibitem{Doplich1}%
S. Doplicher, K. Fredenhagen, J.E. Roberts, {\it{``The quantum structure of spacetime at the Planck scale and quantum fields"}}, Commun. Math. Phys. {\bf{172}} (1995), 187. S. Doplicher, K. Fredenhagen and J. E. Roberts, {\it{``Space-time quantization induced by classical gravity"}}, {Phys. Lett. B\textbf{331}, 39--44 (1994)}. 

\bibitem{majid-ruegg} S. Majid, H. Ruegg, {\it{``Bicrossproduct structure of $\kappa$-Poincar\'e group and non-commutative geometry"}}, Phys. Lett. B{\bf{334}} (1994) 348.

\bibitem{luk1} J. Lukierski, H. Ruegg, A. Nowicki, V. N. Tolsto\"\i, {\it{``$q$-deformation of Poincar\'e algebra"}}, Phys. Lett. B{\bf{264}} (1991) 331. J. Lukierski, A. Nowicki, H. Ruegg, {\it{``New quantum Poincar\'e algebra and $\kappa$-deformed field theory"}}, Phys. Lett. B{\bf{293}} (1992) 344.

\bibitem{leningrad} V. G. Drinfeld, {\it{``Quantum Groups"}}, in Proc. Int. Cong. Math., Vols 1,2 (Berkeley 1986) AMS, Providence, RI (1987) 798. 
L. A. Takhtadzhyan, {\it{``Lectures on quantum groups"}}, Nankai Lectures on Mathematical Physics, Mo-Lin-Ge, Bao-Heng-Zhao Eds., World Scientific (1989).

\bibitem{luk2} J. Lukierski, {\it{``kappa-Deformations: Historical Developments and Recent Results"}}, J. Phys. Conf. Ser. \textbf{804}, 012028 (2017).

\bibitem{ame-ca1} G. Amelino-Camelia, {\it{``Doubly special relativity"}}, Nature {\bf{418}} (2002) 34. G. Amelino-Camelia, G. Gubitosi, A. Marciano, P. Martinetti, F. Mercati, {\it{``A no-pure boost uncertainity principle from spacetime noncommutativity"}}, Phys. Lett. B{\bf{671}} (2009) 298. For a review on Doubly Special Relativity, see e.g J. Kowalski-Glikman, {\it{``Introduction to doubly special relativity"}} in Planck scale Effects in Astrophysics and Cosmology, Lecture Notes in Phys. {\bf{669}} (Springer, Berlin 2005) 131, and references therein.

\bibitem{reloc} G. Amelino-Camelia, L. Freidel, J. Kowalski-Glikman, L. Smolin, {\it{``Principle of relative locality"}}, Phys. Rev. D{\bf{84}} (2011) 084010. G. Gubitosi, F. Mercati, {\it{``Relative locality in $\kappa$-Poincar\'e"}}, Class. Quant. Grav. {\bf{30}} (2013) 145002. G. Amelino-Camelia, V. Astuti, G. Rosati, {\it{``Relative locality in a quantum spacetime and the pregeometry of $\kappa$-Minkowski"}}, Eur. Phys. J. C{\bf{73}} (2013) 2521. 

\bibitem{dnsw-rev}
M.~R. Douglas, N.~A. Nekrasov, {\it{``Noncommutative field theory"}}, {Rev. Mod. Phys. \textbf{73}, 977 (2001)}. R. J. Szabo, {\it{``Quantum field theory on noncommutative spaces"}}, {Phys. Rep. \textbf{378}, 207--299 (2003)}. 
J.-C. Wallet, {\it{``Noncommutative Induced Gauge Theories on Moyal Spaces"}}, {J. Phys. Conf. Ser. \textbf{103}, 012007 (2008)}. %

\bibitem{lagraa} A.~B. Hammou, M. Lagraa, M.~M. Sheikh-Jabbari, {\it{``Coherent state induced star product on R**3(lambda) and the fuzzy sphere"}}, {Phys. Rev. D\textbf{66}, 025025 (2002)}. 

\bibitem{Grosse:2003aj-pc}
H.~Grosse, R.~Wulkenhaar, {\it{``Renormalisation of $\varphi^4$-theory on noncommutative $\mathbb{R}^2$ in the matrix base}}'', JHEP {\bf 0312} (2003) 019. H.~Grosse, R.~Wulkenhaar, {\it{``Renormalisation of $\varphi^4$-theory on noncommutative $\mathbb{R}^4$ in the matrix base}}'', Commun.\ Math.\ Phys.\  {\bf 256} (2005) 305. 

\bibitem{brol-1}H. Grosse, R. Wulkenhaar, {\it{``Self-dual noncommutatif $\varphi^4$-theory in four dimensions is a non-perturbatively solvable and non-trivial quantum field theory"}}, Commun. Math. Phys. {\bf{329}} (2014) 1069. 

\bibitem{vign-sym} F. Vignes-Tourneret, {\it{``Renormalisation of the orientable non-commutative Gross-Neveu model"}}, Ann. H. Poincar\'e {\bf{8}} (2007) 427. 
A. de Goursac, J.-C. Wallet, {\it{``Symmetries of noncommutative scalar field theory}}'', J.\ Phys.\ A: Math. Theor. {\bf{44}} (2011) 055401. 

\bibitem{thes-vt} F. Vignes-Tourneret, {\it{``Renormalisation des th\'eories de champs non commutatives"}}, PhD thesis (2006), [math-ph/0612014] (2006).


\bibitem{vitwal1} P. Vitale, J.-C. Wallet, {\it{``Noncommutative field theories on $\mathbb{R}^3_\lambda$: Toward UV/IR mixing freedom}}'', {JHEP} \textbf{04} (2013) 115.
A. G\'er\'e, P. Vitale, J.-C. Wallet, {\it{``Quantum gauge theories on noncommutative three-dimensional space"}}, {Phys. Rev. D\textbf{90} (2014) 045019}.

\bibitem{poulwal-1}  T. Juri\'c, T. Poulain, J.-C. Wallet, {\it{``Closed star product on noncommutative R3 and scalar field dynamics"}}, JHEP {\bf{05}} (2016) 146. 
T. Juri\'c, T. Poulain, J.-C. Wallet, {\it{``Involutive representations of coordinate algebras and quantum spaces"}}, JHEP {\bf{07}} (2017) 116.

\bibitem{vitwal2}
A. G\'er\'e, T. Juri\'c, J.-C. Wallet, {\it{``Noncommutative gauge theories on $\mathbb{R}^3_\lambda$: Perturbatively finite models"}}, {{JHEP} \textbf{12} (2015) 045}.

\bibitem{wal-16} J.-C. Wallet, {\it{``Exact Partition Functions for Gauge Theories on $\mathbb{R}^3_\lambda$"}}, Nucl. Phys. B{\bf{912}} (2016) 354.

\bibitem{Wallet:2007c}
H.~Grosse, M.~Wohlgenannt, \textit{``Induced gauge theory on a noncommutative space"}, Eur. Phys. J. \textbf{C52} (2007) 435.
A.~de~Goursac, J.-C. Wallet, R.~Wulkenhaar, \textit{``Noncommutative induced gauge theory"}, Eur. Phys. J. \textbf{C51} (2007) 977.

\bibitem{matrix1}
H. Grosse, H. Steinacker, M. Wohlgenannt, {\it{``Emergent Gravity, Matrix Models and UV/IR Mixing"}}, JHEP {\bf{04}}(2008) 023.
A. de Goursac, A. Tanasa, J.-C. Wallet, {\it{``Vacuum configurations for renormalizable non-commutative scalar models}}'', Eur. Phys. J. C{\bf{53}} (2008) 459.
A. de Goursac, J.-C. Wallet, R. Wulkenhaar, {\it{``On the vacuum states for noncommutative gauge theory"}}, {Eur. Phys. J. C\textbf{56} (2008) 293--304}. %
P. Martinetti, P. Vitale, J.-C. Wallet, {\it{``Noncommutative gauge theories on $\mathbb{R}^2_\theta$ as matrix models}}'', {JHEP} \textbf{09} (2013) 051.

\bibitem{matrix2} For a review on emergent gravity in matrix models, see e.g H. Steinacker ,{\it{``Emergent Geometry and Gravity from Matrix Models: an Introduction"}}, Class.Quant.Grav. {\bf{27}}:133001 (2010). 

\bibitem{groen} H.J. Groenewold, {\it{``On the principle of elementary quantum mechanics"}}, Physica {\bf{12}} (1946) 405.

\bibitem{Moyal} J.E. Moyal, {\it{``Quantum Mechanics as a statistical theory"}}, Proc. Cambridge Phil. Soc. {\bf{45}} (1949) 99.

\bibitem{dabrow} L. D\k abrowski, M. Godli\'nski, G. Piacitelli, {\it{``Lorentz covariant $\kappa$-Minkowski spacetime"}}, Phys. Rev. D{\bf{81}}, 125024 (2010). L. D\k abrowski, G. Piacitelli, {\it{``Poincar\'e covariance and $\kappa$-Minkowski spacetime"}}, Phys. Lett. A{\bf{375}} (2011). 

\bibitem{luk3} J. Lukierski, V. Lyakhovsky, M. Mozrzymas, {\it{``$\kappa$-deformations of $D= 4$ Weyl and conformal symmetries}}, Phys. Lett. B{\bf{538}}, 375 (2002).

\bibitem{chaichian} M. Chaichian, P. P. Kulish, K. Nishijima, A. Tureanu, {\it{``On a Lorentz-invariant interpretation of noncommutative space–time and its implications on noncommutative QFT"}}, Phys. Lett. B{\bf{604}} (2004). M. Chaichian, P. Pre\v{s}najder, A. Tureanu, {\it{``New Concept of Relativistic Invariance in Noncommutative Space-Time: Twisted Poincar\'e Symmetry and Its Implications"}}, Phys. Rev. Lett. {\bf{94}} (2005).

\bibitem{GAC2002:2} G. Amelino-Camelia, {\it{``On the fate of Lorentz symmetry in loop quantum gravity and noncommutative spacetimes"}}, [gr-qc/0205125] (2002).

\bibitem{ital-1} A. Agostini, G. Amelino-Camelia, M. Arzano, F. D'Andrea, {\it{``Action functional for kappa-Minkowski non-commutative spacetime"}}, [hep-th/0407227]. A. Agostini, G. Amelino-Camelia, F. D'Andrea, {\it{``Hopf-algebra description of noncommutative-spacetime symmetries"}}, Int. J. Mod. Phys. A{\bf{19}} (2004) 5187. A. Agostini, G. Amelino-Camelia, M. Arzano, A. Marciano, R. Altair Tacchi, {\it{``Generalizing the Noether theorem for Hopf-algebra spacetime symmetries"}}, Mod. Phys. Lett. A{\bf{22}} (2007) 1779.

\bibitem{habsb-imp} M. Dimitrijevi\'c, L. Jonke, L. M\"oller, E. Tsouchnika, J. Wess, M. Wohlgennant {\it{``Deformed field theory on $\kappa$-spacetime"}}, Eur. Phys. J. C{\bf{31}} (2003) 129. M. Dimitrijevi\'c, F. Meyer, L. M\"oller, J. Wess,  {\it{``Gauge theories on the kappa-Minkowski spacetime"}}, Eur. Phys. J. C{\bf{36}} (2004) 117. M. Dimitrijevi\'c, L. Jonke, L. M\"oller, {\it{``$\ U(1)$ gauge field theory on $\kappa$-Minkowski"}}, JHEP {\bf{9}} (2005) 068. M. Dimitrijevi\'c, L. Jonke, A. Pachol, {\it{``Gauge Theory on Twisted κ-Minkowski: Old Problems and Possible Solutions"}}, SIGMA {\bf{10}} (2014) 063.

\bibitem{kappa-star1}
A. Borowiec, A. Pachol, {\it{``kappa-Minkowski spacetime as the result of Jordanian twist deformation"}}, Phys.Rev.D {\bf{79}} (2009) 045012; 
A. Pachol, P. Vitale {\it{``$\kappa$-Minkowski star product in any dimension from symplectic realization"}}, J. Phys. A: Math. Theor. 48 (2015) 445202. S. Meljanac, A. Samsarov, M. Stojic, K. S. Gupta, {\it{``Kappa-Minkowski space-time and the star product realizations"}}, 
Eur.Phys.J.C {\bf{53}} (2008) 295.

\bibitem{hrvat-1} S. Meljanac, A. Samsarov, {\it{``Scalar field theory on kappa-Minkowski spacetime and translation and Lorentz invariance"}}, Int. J. Mod. Phys. A{\bf{26}} (2011) 1439. E. Harikunmar, T. Juri\'c, S. Meljanac, {\it{``Electrodynamics on $\kappa$-Minkowski space-time"}}, Phys. Rev. D{\bf{84}} (2011) 085020. S. Meljanac, A. Samsarov, J. Trampetic, M. Wohlgenannt, {\it{``Scalar field propagation in the $\phi^4$ kappa-Minkowski model"}}, JHEP {\bf{12}} (2011) 010.

\bibitem{gross-whl} H. Grosse, M. Wohlgenannt, {\it{``On $\kappa$-Deformation and UV/IR Mixing"}}, Nucl.Phys. B{\bf{748}} (2006) 473.

\bibitem{star-spunz} M. Dimitrijevi\'c, L. M\"oller, E. Tsouchnika, {\it{``Derivatives, forms and vector fields on the $\kappa$-deformed Euclidean space"}}, J. Phys.A{\bf{37}} (2004) 9749.

\bibitem{ypa} V.  Kathotia, {\it{``Kontsevich's universal formula for deformation quantisation and the Campbell-Baker-Hausdorff  formula"}}, [math.qa/9811174] (1998).

\bibitem{jvn}J. von Neumann, {\it{``Die Eindeutigkeit der Schr\"odingerschen Operatoren"}}, Math. Ann. {\bf{104}}
(1931) 570. See also J. von Neumann ,{\it{``Mathematical foundations of quantum mechanics"}}, Princeton Univ. Press, Princeton, 1955.

\bibitem{DS}  B. Durhuus, A. Sitarz,  {\it{``Star product realizations of kappa-Minkowski space"}}, J. Noncommut. Geom. \textbf{7} (2013) 605.

\bibitem{matas} M. Matassa, {\it{``On the spectral and homological dimension of k-Minkowski space"}}, [arXiv:1309.1054] (2013). M. Matassa, {\it{``A modular spectral triple for $\kappa$-Minkowski space"}}, J. Geom. Phys. {\bf{76}} (2014) 136.

\bibitem{dana} D. Williams, {\it{``Crossed Products of C*-algebras"}}, Math. Surveys and Monographs, Vol. 134, AMS (2007).

\bibitem{ypa-2} A.L. Onishchik, E.B. Vinberg (Eds.), {\it{``Lie groups and Lie algebras III : structure of Lie groups and Lie algebras"}}, EMS Vol. 41, Springer,  1994.

\bibitem{khalil} I. Khalil, {\it{``Sur l'analyse harmonique du groupe affine de la droite"}}, Studia Math., {\bf{51}} (1974) 139.

\bibitem{vanbrunt} A. Van-Brunt, M. Visser, {\it{``Special-case closed form of the Baker-Campbell-Hausdorff formula"}}, 
 J.\ Phys.\ A: Math. Theor. {\bf{48}} (2015) 225207.

\bibitem{hennings} M. A. Hennings, D. A. Dubin, {\it{``On Hansen's Version of Spectral Theory and the Moyal Product"}}, Publ. RIMS, Kyoto Univ. {\bf{45}} (2009) 1041.

\bibitem{kuster} J. Kustermans, {\it{``KMS-weights on C*-algebras"}}, [funct-an/9704008] (1997). See also F. Combes
, {\it{``Poids sur une C*-alg\`ebre"}}, J. Math. pures et appl. {\bf{47}} (1968) 57. 

\bibitem{como-1} A. Connes, H. Moscovici, {\it{``Type III and spectral triples"}}, in Traces in number theory,geometry and quantum fields, Aspects of Math. E38, Vieweg, Wiesbaden 2008, pp 57.

\bibitem{takesaki} M. Takesaki, {\it{``Theory of Operator Algebras I-III"}}, EMS Vols. 124, 125, 127, Springer 2002.

\bibitem{ConRove} A. Connes, C. Rovelli, {\it{``Von Neumann Algebra Automorphisms and Time-Thermodynamics Relation in General Covariant Quantum Theories"}}, Class.Quant.Grav. {\bf{11}} (1994) 2899. 

\bibitem{frans-2} F. D'Andrea, {\it{``Spectral geometry of $\kappa$-Minkowski space"}}, J. Math. Phys. {\bf{47}} (2006) 062105.

\bibitem{GAC2002} G. Amelino-Camelia, M. Arzano, {\it{``Coproduct and star product in field theories on Lie-algebra noncommutative space-times"}}, {Phys. Rev. D {\bf{65}}} (2002) 084044.

\bibitem{PW-1} T. Poulain, J.-C. Wallet, {\it{``Beta-functions for 4 dimensional $\kappa$-Poincar\'e invariant field theories"}}, in preparation.

\bibitem{PSW-1} T. Poulain, R. \v{S}trajn, J.-C. Wallet, {\it{``Quantum behavior of gauges theories on $\kappa$-Minkowski"}}, in preparation.

\bibitem{Maj-sit} S. Majid, {\it{``Classification of bicovariant differential calculi"}},  J. Geom. Phys. {\bf{25}} (1998) 119. See also A. Sitarz, {\it{``Noncommutative differential calculus on the $\kappa$-Minkowski space"}}, Phys. Lett. B{\bf{349}} (1995) 42.

\bibitem{mdv-jcw} M. Dubois-Violette, {\it{``Lectures on graded differential algebras and noncommutative geometry"}}, Noncommutative Differential Geometry and Its Applications to Physics, Springer Netherlands, 245-306 (2001). J.-C. Wallet, {\it{``Derivations of the Moyal algebra and Noncommutative gauge theories"}}, {SIGMA \textbf{5} (2009) 013}. 
E. Cagnache, T. Masson, J-C. Wallet, {\it{``Noncommutative Yang-Mills-Higgs actions from derivation based differential calculus"}}, {J. Noncommut. Geom. \textbf{5}, 39-67 (2011)}. 
A.~de~Goursac, T.~Masson, J.-C. Wallet, \textit{{``Noncommutative $\varepsilon$-graded connections"}}, {J. Noncommut. Geom.} \textbf{6} (2012) 343-387.
\end{thebibliography}
\end{document}